\documentclass{emulateapj}
\pdfoutput=1
\usepackage{apjfonts,graphicx,amsmath,amssymb,amsfonts,natbib,microtype}

\newcommand{\Ntod}{N_{\text{tod}}}
\newcommand{\Npix}{N_{\text{pix}}}

\newcommand{\bigoh}{{\mathcal O}}

\newcommand{\like}{\ensuremath{\mathcal{L}}}
\newcommand{\vect}[1]{\ensuremath{\mathbf{#1}}}

\begin{document}
\submitted{Submitted to ApJ}
\shorttitle{NEW CMB POLARIZATION RESULTS FROM CAPMAP}
\shortauthors{CAPMAP COLLABORATION}
\title{New Measurements of Fine-Scale CMB Polarization Power \\ Spectra from CAPMAP at Both 40 and 90~GHz}
\author{CAPMAP~Collaboration
--
C.~Bischoff,\altaffilmark{1}
L.~Hyatt,\altaffilmark{2}
J.~J.~McMahon,\altaffilmark{2,3}
G.~W.~Nixon,\altaffilmark{2}
D.~Samtleben,\altaffilmark{4}
K.~M.~Smith,\altaffilmark{1,5}
K.~Vanderlinde,\altaffilmark{1,6}
D.~Barkats,\altaffilmark{2,7}
P.~Farese,\altaffilmark{2,8}
T.~Gaier,\altaffilmark{9}
J.~O.~Gundersen,\altaffilmark{10}
M.~M.~Hedman,\altaffilmark{1,11}
S.~T.~Staggs,\altaffilmark{2}
and
B.~Winstein\altaffilmark{1}}

\altaffiltext{1}{Kavli Institute for Cosmological Physics, Department of Physics, Enrico Fermi Institute, The University of Chicago, Chicago, IL 60637.}
\altaffiltext{2}{Department of Physics, Princeton University, Princeton,
NJ 08544.}
\altaffiltext{3}{Currently an Enrico Fermi Fellow, University of Chicago.}
\altaffiltext{4}{Max-Planck-Institut f\"{u}r Radioastronomie, D-53121
Bonn, Germany.}
\altaffiltext{5}{Currently a PPARC Postdoctoral Fellow, Institute of Astronomy, University of
Cambridge, Madingley Road, Cambridge, CB3 0HA, United Kingdom.}
\altaffiltext{6}{Current address: South Pole Telescope, Project A-379, South Pole Station, PSC 468 Box 400, APO AP 96598.}
\altaffiltext{7}{Current address: Department of Physics, California
Institute of Technology, Pasadena, CA 91125.}
\altaffiltext{8}{Current address: McKinsey \& Co, 3 Landmark Sq, Ste 100, Stamford, CT 06901.}
\altaffiltext{9}{Jet Propulsion Laboratory, California Institute of
Technology, Oak Grove Drive, Pasadena, CA 91109.}
\altaffiltext{10}{Department of Physics, University of Miami, Coral
Gables, FL 33146.}
\altaffiltext{11}{Current address: Department of Astronomy, Space Sciences Bldg, Cornell University, Ithaca, NY 14853.}

\begin{abstract}
We present new measurements of the cosmic microwave background (CMB) polarization from the  final season of the Cosmic Anisotropy Polarization MAPper (CAPMAP).  The data set was obtained in winter 2004--2005 with the 7~m antenna in Crawford Hill,  New Jersey, from 12 W-band (84--100~GHz) and 4 Q-band (36--45~GHz) correlation polarimeters with $3.3\arcmin$ and $6.5\arcmin$ beamsizes, respectively.  After selection criteria were applied,
956 (939) hours of data survived for analysis of W-band (Q-band) data.
 Two independent and complementary pipelines produced results in excellent agreement with each other.  A broad suite of null tests as well as extensive simulations showed that systematic errors were minimal, and a comparison of the W-band and Q-band sky maps revealed no contamination from galactic foregrounds.  We report the $E$-mode and $B$-mode power spectra in 7 bands in the range $200\lesssim\ell\lesssim3000$, extending the range of previous measurements to higher $\ell$. The $E$-mode spectrum, which is detected at $11\sigma$ significance, is in agreement with cosmological 
predictions and with previous work at other frequencies and angular resolutions. The $BB$ power spectrum provides one of the best limits to date on $B$-mode power at 4.8~$\mu$K$^2$ (95\% confidence).
\end{abstract}


\section{Introduction} 
The CMB has not yet yielded all its secrets, particularly from its polarization. Although the outlines of the power spectrum of the positive-parity $E$-modes have emerged since their first detection five years ago \citep{Kovac:2002}, its details have not, and the negative-parity $B$-modes remain entirely elusive.  The principle sources of $E$-modes are the same perturbations in the primeval plasma that gave rise to temperature anisotropies. Since the latter are well-measured, robust predictions exist for the cosmological $EE$ spectrum on most angular scales.  

To date, $EE$ data from some seven collaborations are in agreement with predictions. Nonetheless, more detailed characterization of the $EE$ spectrum, one of the very few probes of the early universe, is important and may reveal surprises. Additionally, these efforts lay groundwork for future detection of  $B$-modes, which are compelling targets.  At the subdegree angular scales studied by CAPMAP, $B$-modes should arise from gravitational lensing of $E$-modes, and thus provide a new channel for  examining the intervening energy content of the universe.  At large angular scales their sole extragalactic sources are gravitational waves from the earliest instants of time; detection of those primordial $B$-modes would give a measure of the energy scale of inflation itself.  

Here we report further measurements of the $EE$ spectrum and new constraints on the $B$-mode  power from the final season of CAPMAP.  These data derive from a unique combination of observing frequencies and angular resolution, and in particular extend to larger multipoles than previous work.
CAPMAP measured the CMB polarization at 40~GHz and 90~GHz simultaneously, two frequencies chosen to bracket the foreground frequency minimum observed by WMAP and to permit new constraints on foreground contaminants. Moreover, the measurements were made at very small angular scales using correlation polarimeters rather than bolometers or interferometers.  Thus, we report here data complementary to that from recent work (including our own): \citet{quad,Barkats:2004he, dasi, boomerang, wmappol, cbi, maxipol}. 
 
The reported power spectra, drawn from two complementary analysis pipelines, are robust, as evidenced by results from a suite of 72 null tests on the polarization maps, and from extensive simulations of possible systematic effects.  

We begin with description of the experimental configuration (\S\ref{sec:instrument}), the CMB observations (\S\ref{sec:cmbobs}), and the methods of calibration  (\S\ref{sec:calib}), before moving on to explanation of the data analysis (\S\ref{sec:analysis}). \S\ref{sec:results} presents the polarization power spectra and the results of the null tests, while \S\ref{sec:systematics} and \S\ref{sec:foregrounds} discuss evaluation of systematic errors and foreground limits. We finish with comments on internal consistency (\S\ref{sec:comparisons}) followed by conclusions (\S\ref{sec:conclusions}).

\section{Instrument Description}\label{sec:instrument}
\begin{figure}
\centering
\includegraphics[width=3.39375in]{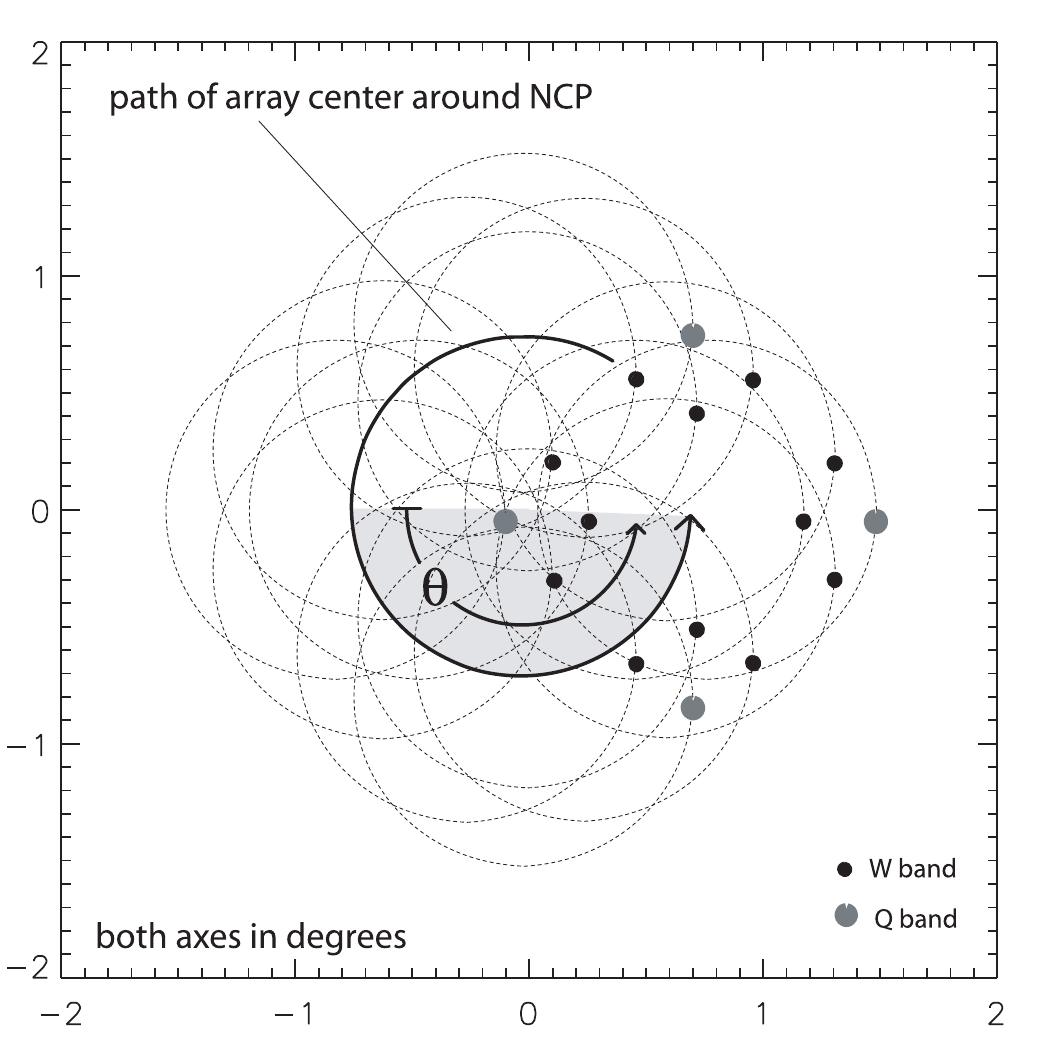}
\caption{\label{fig:scan}
CAPMAP array configuration and the ring scan.   The instantaneous locations of the 16 radiometers are shown with black (gray) circles for W-band (Q-band). The circle diameters show the $3.3^\prime$ ($6.5^\prime$) beams.  The telescope moved so that the center of the array continuously traced a constant-declination circle about the NCP. Dashed circles indicate sky coverage: each is the path followed by one radiometer during a single 21~s cycle. Each radiometer timestream is considered to be a function of the variable $\theta$.}
\end{figure}

The CAPMAP instrument is described in \citet{Barkats:2005} and summarized in \citet{Barkats:2004he}. Below we outline the final configuration used for the observations reported herein, highlighting improvements compared to the first season. The CAPMAP instrument was optimized to measure the $EE$ and $BB$ power spectra, sacrificing sensitivity to $TT$ and $TE$.

CAPMAP comprised 16 heterodyne correlation polarimeters coupled via corrugated feed horns and high-density polyethylene lenses to the 7~m Cassegrain antenna \citep{Chu:1978} in Crawford Hill, NJ (W$74^\circ 11^\prime 11^{\prime\prime}$, N$40^\circ 23^\prime 31^{\prime \prime})$. 12 receivers operated at W-band (84--100~GHz) with beam
FWHM $3.3^\prime$, while the other four operated at Q-band (35--46~GHz) with beam FWHM  $6.5^\prime$. The receivers were organized into four groups (numbered 0 to 3), each including one Q-band and three W-band receivers (labeled B, C and D) housed in a cryostat, which cooled the high-electron-mobility transistor low-noise amplifers (LNAs) and feedhorns to 25~K, and the lenses to 80~K. 

For each radiometer, the circular feed coupled two orthogonal components of the incident electric field via an orthomode transducer into LNAs in each arm.  The LNAs, provided by JPL, were based on monolithic microwave integrated circuits (MMICs) and had typical noise temperatures of 65~K for W-band and 35~K for Q-band. After amplification and bandpass filtering, mixers downconverted the radio frequency signals to intermediate frequencies (IF)  in the range 2--18~GHz with  local oscillators (LOs) operating at 82 GHz or 30.5 GHz. The LO signal for one arm of each polarimeter was phase-switched at 4~kHz to suppress 1/$f$ noise. The three W-band radiometers in each cryostat shared a single LO.

After IF amplification, the W-band (Q-band) signals were divided into three (two) frequency subbands per polarimeter, for a total of 44 channels.  For each, signals from the two arms were combined in an analog multiplier and then preamplified before digitization. Thus each polarimeter timestream was proportional to the product of the two orthogonal modes of the incoming electric field. At a given parallactic angle $\psi$, each CAPMAP polarimeter measured one linear combination of the Stokes parameters $Q$ and $U$,
\begin{equation}
	P = Q \cos \bigl[2(\psi + \eta)\bigr] + U \sin \bigl[2(\psi + \eta)\bigr],
\end{equation}
where the angle $\eta$, the \emph{detector polarization angle}, varied for each channel but was nominally $45^\circ$. 

The subbands were labeled S0, S1 and (for W-band) S2. The two orthogonally-polarized total power signals were tapped off each arm of a polarimeter prior to the frequency subdivision, and  recorded with detector diodes.  Because they were not phase-switched, these total power channels were too noisy to provide CMB data, but they were useful for assessing radiometer health, monitoring the atmospheric noise for data selection, measuring the properties of the beams, and determining the atmospheric opacity.

The active components (the cold LNAs, the warm LOs, the IF amplifiers and the pre-amplifiers) were temperature-controlled to stabilize the responses of the radiometers. These servo systems did not perform consistently, however.  In particular, their temperature set-points had to be increased multiple times, resulting in changes of 10--20~K from the start to the end of the observing season.
Moreover, the temperatures occasionally drifted for 5--10 hour stretches in response to large swings in the ambient temperature. These drifts were typically at the level of 0.3 K~hr$^{-1}$ but reached as high as 1--2 K~hr$^{-1}$ in some cases.
As described in \S\ref{sec:systematics}, these temperature instabilities led to a systematic error in the experiment, but it was considerably smaller than the statistical error.

Both the polarization and total power channels were sampled at 100 kHz by a commercial sigma-delta ADC board, as described in \citet{Barkats:2005}. In lieu of recording high-speed data as in the first observing season of CAPMAP,  the polarization channels were digitally demodulated both in phase and $90^\circ$ out of phase with the 4~kHz phase switch clock. The out-of-phase (quadrature) data lacked signal from the sky, but had the same white noise properties as  the in-phase data and so were used for systematic checks. A small amount of signal (typically $<2\%$)  did leak into the quadrature data, due to phase shifts in the data relative to the digitized clock used for demodulation, but this led to a negligible sensitivity reduction for the in-phase data. The data rate was 2.9~GB per day. 

\begin{figure}[b]
\centering
\includegraphics[width=3.39375in]{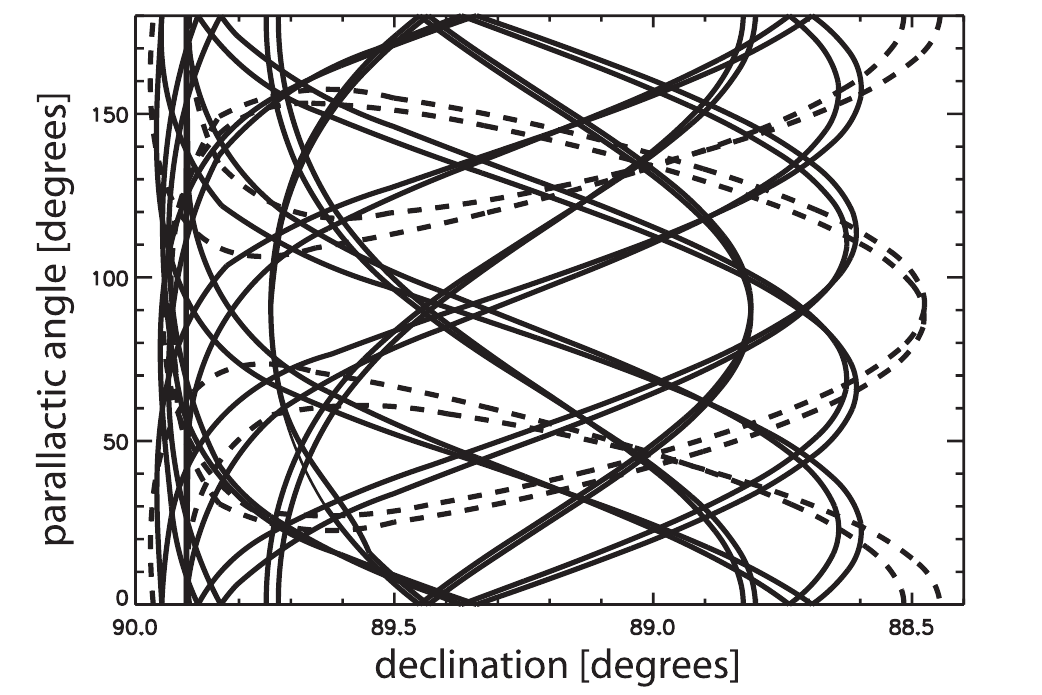}
\caption{\label{fig:coverage} Parallactic angle coverage.  The solid (dashed) lines show the paths of the 12 W-band (4 Q-band) radiometers during a single 21~s cycle. Pixels at each declination were observed at a variety of parallactic angles, overconstraining the $Q$ and $U$ Stokes parameters everywhere in the map.}
\end{figure}

The array configuration and scan strategy, illustrated in Figure~\ref{fig:scan}, were designed to have high symmetry, provide highly cross-linked maps,
and constrain $Q$ and $U$ tightly with uniform parallactic angle coverage. For CMB observations, the telescope's optical axis described a circle around the north celestial pole (NCP) with a radius $r=0.725^\circ$ and a period of 21~s. The observed position is parameterized by the ring variable $\theta$ shown in Figure~\ref{fig:scan}. This scanning method surveyed a $3^\circ$ disc centered on the NCP, covering the entire region every six hours with good noise uniformity ($\pm50\%$) and an even parallactic angle distribution (Fig.~\ref{fig:coverage}). The 7~m antenna provided a large and uniform focal plane so that even the outermost receivers had essentially Gaussian main beam profiles.

The fact that the elevation varied continuously throughout the scan provided several side benefits. The ring motion produced a $\sin\theta$ modulation in the total power channels with amplitude given by \citep{jeff_thesis}
\begin{equation}\label{eq:opacity}
    A_1 = - \tau T_{\text{atm}}\cot(el_0)\csc(el_0)r\exp\bigl[-\tau \csc(el_0)\bigr],
\end{equation}
where $\tau$ is the opacity of the atmosphere, $T_{\text{atm}}$ is its temperature, $el_0= 40.3^\circ$ is the central elevation of the ring scan, and $r$ its radius. The total power channels thus continuously monitored the opacity and the stability of the atmosphere. Since the atmospheric signal is an unpolarized intensity $I$, sinusoidal responses in the polarimetry channels measured the small leakage of $I$ into polarization, which we refer to as $I \rightarrow Q$ leakage.  As discussed further in \S\ref{sec:data_selection} and \S\ref{sec:mapmaking}, the atmospheric signal can be cleanly removed from the polarization channels (with a 15\% loss of sensitivity).

We note here the significant improvements in the final instrument configuration for CAPMAP since the first season (reported in  \citet{Barkats:2004he}).  The original four W-band receivers have been supplemented by an additional eight W-band and four  Q-band receivers.  Data were collected with the ring scan strategy described above, rather than with a constant-elevation scan.  The lenses were  upgraded to reduce $I\rightarrow Q$ leakage, primarily by improving their anti-reflection (AR) coating.  The original coatings comprised  birefringent grooves, which were replaced with a polarization-symmetric square grid of round holes; the resulting improvement is reported in  \S\ref{sec:IQleak}.  
Finally, an order-of-magnitude reduction of scan synchronous signals from telescope sidelobes was accomplished in a four-pronged approach \citep{jeff_thesis}:  1) a co-moving ground screen was added between the receivers and the secondary mirror;  2) a baffle was added covering the edges of the secondary mirror, redirecting rays otherwise headed for the ground back onto the primary and thence to the sky; 
3) gaps between panels of the primary mirror were sealed with copper tape, measurably reducing diffractive sidelobes from those gaps; and 4) several nearby trees that had previously illuminated sidelobes were removed. 

These modifications reduced the linear dependence of the polarimeter response on azimuth  by an order of magnitude, from 500~$\mu$K~deg$^{-1}$  to  50~$\mu$K~deg$^{-1}$.  Moreover, whereas these slopes showed variations of  $\approx 200$~$\mu$K~deg$^{-1}$~day$^{-1}$ during the first season,  any remaining variations were  no longer measurable.
Concluding remarks on the efficacy of this sidelobe-reduction effort are given in \S\ref{sec:data_selection} and \S\ref{sec:results}.

\section{CMB Observations}\label{sec:cmbobs}
Between 13 December 2004 and 28  April  2005, CAPMAP recorded 1658 hours of CMB data: a 53\% livetime.
Another 3\% of the time was devoted to calibration data.  The majority of the downtime was due to snow, rain, and heavy clouds, but power outages, telescope repair, and other gross failures prevented observations about 10\% of the time.  We downselected the CMB data as described in \S\ref{sec:data_selection}, primarily removing more periods of bad weather.  The final sample comprised data from 29\% of the time in the interval above.

We divided the data into three periods (I, II, and III) of roughly equal length and sensitivity, separated by power outages.  These periods are visible in Figures~\ref{fig:tau}~and~\ref{fig:gains}.  For the purposes of data selection and quality inspection, we further subdivided the data into nine subperiods.  We also designated one long continuous segment of one of the subperiods as the reference period (RP), used for initial exploration of the data.   

\section{Calibration}\label{sec:calib}
In addition to CMB data, calibration data were taken before, during, and after the season.  These data measured parameters necessary for translating the CMB timestreams into polarization power spectra, including the coefficients in the pointing model, the beam sizes and shapes, the detector polarization angles, and the calibration coefficients $C$ relating the detector voltage outputs to thermal CMB fluctuations.  These last require measurement of both the atmospheric opacity $\tau$ and the detector responsivities $\cal R$ (with units V~K$^{-1}$) to thermal sources below the atmosphere:
\begin{equation}\label{eq:respons}
C = {\cal R}/\exp[\tau\csc(el)],
\end{equation}
where $el$ is the elevation of the observation.  We refer to the denominator in equation~\eqref{eq:respons} as the \emph{absorption correction}. The primary calibration method for $\mathcal{R}$ required evaluation of telescope beam efficiencies. We also used the calibration measurements to determine the level of $I \rightarrow Q$ leakage.

Below we describe the measurements of all these quantities.  We conclude with a description of the detector responsivity measurements, and the model used to extrapolate those measurements to all times during the observing season.  The impact of uncertainties in determining all these parameters is assessed in \S\ref{sec:systematics}. 

\subsection{Pointing Model}\label{sec:pointing}
To determine where the receivers were pointed during CMB and calibration observations, we undertook measurements of both the global pointing of the array center, and the relative offsets of each radiometer from a fixed position in the focal plane.

The global pointing relied on a 10-parameter pointing model \citep{meeks}, constrained by two sets of radio data: one obtained on 31 Dec 2003 with 154 source observations, and another over 26--27 Nov 2004 with 50 observations. In each session, a single Q-band receiver was repeatedly rastered over a succession of five sources (Cas~A, Cyg~A, Tau~A, OMC~1, and Jupiter) as the sky rotated.  Each source map was fit to a Gaussian beam shape, and the recovered center in telescope coordinates was compared to the known source position.  The resulting pointing offsets were used to fit the pointing model parameters.  The residuals from the best fit had an RMS of  $29\arcsec$ in elevation and $18\arcsec$ in cross-elevation.  The results from both data sets were statistically consistent, showing the pointing solution to be stable in time.   

The relative offsets of the radiometers were determined from 11 observations on 26--27 Nov 2004,  in which the full array was rastered across Jupiter.  Again, source maps were fit to Gaussian beams.  The measured offsets agreed well with optical simulations, to within the known alignment tolerances.  The uncertainties in the offsets were typically $4\arcsec$ and always smaller than $7\arcsec$.

\subsection{Beam Measurements}\label{sec:beams}
The beam size and shape for each radiometer were measured with observations of Jupiter in the total power channels.  Two scan strategies were used, both optimized for efficiency by concentrating integration time on single receivers.  The first strategy consisted of a tight raster scan with a $21\arcmin$ width and a $0.3\arcmin$ elevation step. The second, a \emph{point and integrate} scan, used discrete steps spaced on a $1.5^\prime$ grid with a 2~s integration time at each point, bracketed in time by 2~s off-source measurements. For both scan types, the resulting maps were fit to two-dimensional Gaussians to extract the beam parameters.

The two scan strategies produced consistent estimates of the mean beam FWHM in each frequency channel: $3.3^\prime$ and $6.5^\prime$ for W- and Q-band respectively. The standard deviation of the beam size measurements of all W-band (Q-band) radiometers was 2\% (3\%);  all were within $5\%$ of the mean in each case.
We note that the 2Q beam---the highest in the focal plane---was slightly truncated as it left the vertex cab, which elongated it and increased its beam size by about 6\% compared to those from the other three Q radiometers.
 
Since the center frequencies of the polarimeter subbands differed from those of the broadband total power channels, and since the optics were near diffraction-limited, the beam sizes of the high and low subbands differed by about $\pm 2\%$ from the measured total power values \citep{jeff_thesis}.
 
The beams were all quite close to circular, with elongations  less than 8\%.  Elongation is defined as $(b - a) / (b+a)$, where $a$ and $b$ are the beam radii along the principal axes.

\subsection{$I \rightarrow Q$ Leakage}\label{sec:IQleak}
Jupiter measurements also constrained the  $I \rightarrow Q$ leakage parameters of each receiver; the point and integrate scan was optimized specifically for this purpose.  Since the $1/f$ noise of the polarization channels was negligible over the four seconds per point in the map, it was possible to integrate down to measure the very small polarized signals due to optical and radiometric $I \rightarrow Q$.  These measurements showed that the monopole, dipole, and quadrupole terms (see \citet{HHZ:2003} or \citet{Barkats:2005} for the definitions) were below $-23$ dB, $-15$ dB, and $-20$ dB respectively. These levels are small enough to neglect, as discussed in \S\ref{sec:systematics}.  
The new lens AR coatings described in \S\ref{sec:instrument} improved the quadrupole suppression by a factor of ten compared to the original grooved coatings used during CAPMAP's first season.

The monopole leakage terms were also measured by comparing each polarimeter's response to the atmospheric $\sin\theta$ variation in the CMB ring scan to the response from the total power channels, as mentioned in \S\ref{sec:instrument}. These measurements were consistent with those from Jupiter.

\subsection{Detector Polarization Angle Measurements}\label{sec:polznang}
\begin{figure}
\centering
\includegraphics[width=2.4in]{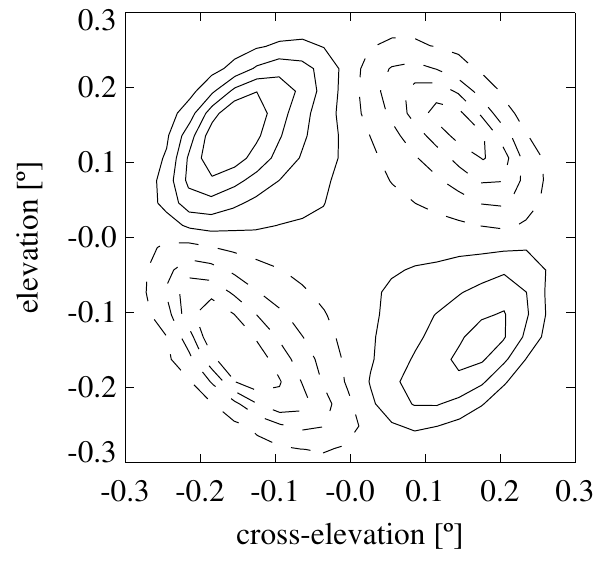}
\caption{\label{fig:moon}Polarized map of the moon from channel 2BS0. The dashed contours indicate negative signal levels. The polarization angle for this channel was fitted to be $44.7^\circ \pm 1.5^\circ$. The moon was not full, as can be seen by the envelope on the quadrupole pattern, which features deeper lobes on the left side; this effect was taken into account when determining the detector polarization angles.}
\end{figure}
Observations of the moon were used to determine the detector polarization angles $\eta$. Thermal radiation from the moon is polarized according to its propagation direction relative to the normal to the surface (see \S\ref{sec:gains} for a related discussion). Thus, the polarization pattern of the moon is oriented radially and peaks in amplitude near the edge of the moon's disc. Projected onto the sensitivity axes of a CAPMAP polarimeter, the radial pattern was seen as a quadrupole, which we fitted to find the angle $\eta$ as illustrated in Figure~\ref{fig:moon}. The measurement error on $\eta$ ranged from $1^\circ$ to $2.5^\circ$; the resulting systematic uncertainty in the final power spectra was small, as discussed in \S\ref{sec:systematics}.

These measurements of $\eta$ also revealed that one radiometer was installed nearly $20^\circ$ rotated from its design orientation, which was confirmed during decommissioning. This affected the parallactic angle coverage but was taken into account in the analysis.

Observations of Tau A at multiple parallactic angles were also fitted for $\eta$ and gave consistent results. 

\subsection{Opacity Determination}\label{sec:opacity}
\begin{figure*}
\centering
\includegraphics[width=\textwidth]{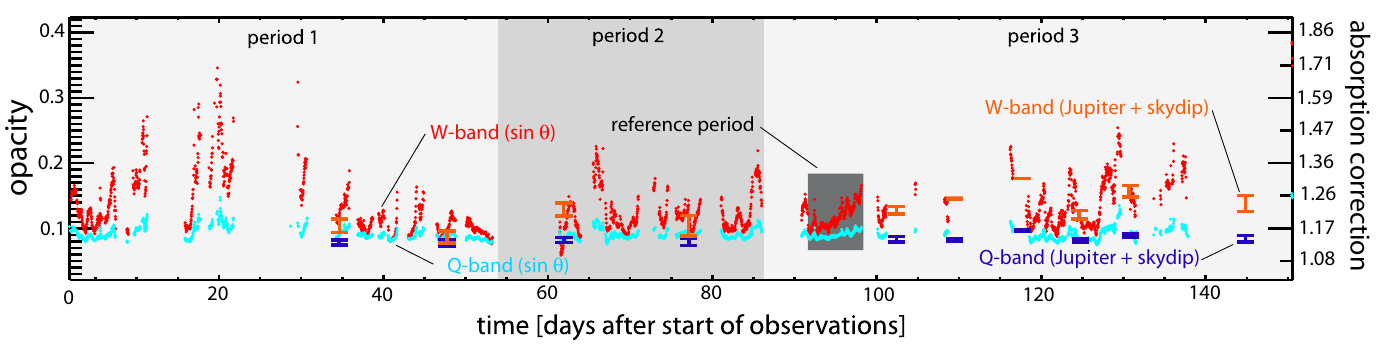}
\caption{\label{fig:tau}The atmospheric opacity (left axis) and the corresponding absorption correction (right axis) for all retained data as a function of time since the start of the observing season. Shown are the separate W- and Q-band measurements, from both the radiometer data (labeled $\sin\theta$) and the 10 Jupiter plus sky dip observations.  Shadings indicate the three periods and the reference period discussed in \S\ref{sec:cmbobs}.}
\end{figure*}

The ring scan permitted continuous monitoring of the opacity during CMB observations. Ten times during the season, the opacity was independently measured from Jupiter scans followed by sky dips, in which the telescope was scanned in elevation from the zenith to near the horizon and back.  The combination of the Jupiter and sky dip data were analyzed to yield both total power responsivity estimates and opacity estimates.  These opacity measurements confirmed the continuous monitoring results, as shown in Figure~\ref{fig:tau}.

The clear sky brightness temperature data from the Geostationary Operational Environmental Satellite\footnote{\url{http://cimss.ssec.wisc.edu/goes/}} were used to estimate $T_{\text{atm}}$, which varied from 255--295~K over the course of the season.  The amplitude $A_1$ (eq.~\eqref{eq:opacity}) was averaged each hour across the 7 (6) most stable W-band (Q-band) total power channels.
Numerical evaluation of equation~\eqref{eq:opacity} resulted in the hourly estimates of $\tau$ plotted in Figure~\ref{fig:tau},  which were splined to the 21~s intervals on which the calibration was applied.
The average value for $\tau$ after data selection was 0.135 (0.092) for W-band (Q-band), corresponding to an absorption correction of 1.23 (1.15).

Measurements of $\tau$ were limited to 10\% accuracy by uncertainties in the total power responsivities used to convert $A_1$ to temperature units. The resulting uncertainty in the absorption correction depended on $\tau$, but was in the worst case 5\% (2\%) for W-band (Q-band).  
The season average was a 1.7\% (1.6\%) uncertainty in the absolute calibration for W-band (Q-band); the impact of this uncertainty is discussed in \S\ref{sec:systematics}.

These $\tau$ measurements agreed with the Jupiter plus sky dip measurements at the 5\% (7\%) RMS level for W-band (Q-band), consistent with errors in the latter.

\subsection{Detector Responsivity Measurements}\label{sec:chopper}
As in the CAPMAP first season, the primary measurement of the responsivity ${\cal R}$  in V~K$^{-1}$ for each polarimeter was made using a nutating aluminum flat (the chopper plate)  installed in front of the secondary mirror \citep{Barkats:2004he,Barkats:2005}.  When radiation from the sky is reflected obliquely by the plate, the non-zero resistivity of aluminum leads to different reflection coefficients for the transverse electric and transverse magnetic boundary conditions, producing a polarized signal from initially unpolarized radiation. The same effect also causes the thermal radiation emitted from the plate to be polarized; the net signal incident on a polarimeter is given in temperature units by
\begin{equation}\label{eq:chopP}
P(\alpha) = \left[\frac{(e^x-1)^2}{x^2e^x}\right]4\sqrt{\pi\epsilon_0\rho\nu}\,Y(T_{\text{sky}}-T_{\text{plate}})\alpha,
\end{equation}
where $\alpha$ is the nutation angle of the plate, $\nu$ is the central (band-averaged) frequency, $\rho$ is the resistivity of the plate,  $Y$ is a geometrical factor of order unity, $T_{\text{plate}}$ is the temperature of the aluminum, and $T_{\text{sky}}$ is the sum of the atmospheric temperature and the CMB temperature, $T_{\text{CMB}}$. The quantity in brackets involving $x=h\nu/kT_{\text{CMB}}$ is the \emph{thermodynamic correction}, which puts the responsivity into  temperature units appropriate for fluctuations in a blackbody source at $T_{\text{CMB}}$. The full nutation range is $\Delta\alpha=13^\circ$, producing a modulated signal with peak-to-peak amplitude $\Delta P$ approximately 90~mK for typical sky and plate temperatures. The responsivity is found from  
${\cal R} =  \epsilon_m \Delta V/\Delta P$, where $\epsilon_m$ is the beam efficiency  \citep{kraus}, described in the following paragraphs, and $\Delta V$ is the measured amplitude from the polarimeter in volts. The uncertainty on $\Delta P$ is $7.5\%$, dominated by a 5\% uncertainty on the value of $\sqrt{\rho}$, and $4\%$ uncertainties on each of $\Delta\alpha$ and $(T_{\text{sky}}-T_{\text{plate}})$.

Because the chopper plate intercepted the beam before the primary and secondary mirrors, each responsivity measurement from it had to be corrected 
to account for the telescope beam efficiency $\epsilon_m$. 
A random surface error RMS of $\sigma_s$ decreases the responsivity to a beam-filling thermal source by the Ruze factor \citep{ruze} at the wavelength $\lambda$: 
\begin{equation}\label{eq:ruze}
\epsilon_m= \exp \left[ - \left ( \frac{ 4\pi\sigma_s}{\lambda} \right)^2 \right]. \end{equation}  
We estimated $\epsilon_m$ and an associated uncertainty for each frequency channel using simulations constrained by measurements.  The measurements included observations of the moon with the CAPMAP receivers, and original surface measurements from \citet{Chu:1978}.  
Specifically, \citet{Chu:1978} reported the range of measured RMS on the individual panels of the primary mirror before installation ($\sigma_p$) and the existence of a 150--200~$\mu$m step ($d$) between the top and bottom halves of the mirror. Furthermore, they measured $\epsilon_m$ at 100~GHz with a prime-focus feed system and inferred $\sigma_s = 100~\mu$m with a $3\sigma$ error of 30~$\mu$m. The secondary mirror surface errors were negligible.

The simulations treated the phase error on a $2048 \times 2048$ grid  covering the telescope aperture.  Each simulation drew $\sigma_p$ for each panel from a Gaussian distribution $\sigma_p = 40\pm 20~\mu$m, and $d$ from $175 \pm 25~\mu$m.  Moreover, each panel was considered to have an additional fourth-order polynomial distortion, consistent with its having an adjuster in each corner; these distortions were constrained by the requirement that the overall surface RMS for all the gridded points be drawn from a flat distribution between 70--130~$\mu$m.  In \S\ref{sec:systematics} we describe a method, using the sidelobes from these simulations, to estimate the systematic effect from ignoring them in the analysis.

The simulations produced beam patterns consistent with the 100-GHz one plotted in \citet{Chu:1978} when the appropriate feed pattern was used; that feed resulted in a 3~dB illumination contour that encompassed the central eight of the 27 panels on the mirror. By contrast, the CAPMAP W-band receivers each illuminated approximately two panels at that level.  For each W-band receiver, many realizations of the simulated beam patterns were each convolved with a $0.5^\circ$ thermal source and compared to 
the moon data, which comprised measurements of the total power response out to $1^\circ$ from the moon center.  Good qualitative agreement was obtained, but only for simulations which conformed to the requirement that $\sigma_s$ was contained in $[70,130]~\mu$m.  From consideration of simulations of all the W-band receivers, we estimated $\epsilon_m = 0.89 \pm 0.05$.  We note that this estimate is in good agreement with the value $0.87\pm 0.03$ obtained from inserting $\sigma_s = 100\pm 10~\mu$m into equation~\eqref{eq:ruze}. 

For Q-band, similar simulations were made but studied in less detail since the final estimated beam efficiency was near unity:  $\epsilon_m = 0.97 \pm 0.01$.  The final absolute calibration errors from combining the $\Delta P$ and $\epsilon_m$ uncertainties in quadrature were 9.4\% (7.6\%) for W-band (Q-band).  

The responsivity for each of the 44 polarization channels was determined from chopper plate observations at 10 separate times, spaced approximately evenly throughout the observing season. In each of the 10 measurements, the chopper plate was run for approximately 30 min and the amplitude was fit in 6~s intervals containing $\approx3$ nutation periods each, determining the responsivities with better than $1\%$ statistical precision. 

The responsivity of each channel was also measured through 6--7 observations of the polarized source Tau~A during and after the season. The absolute accuracy of these measurements was limited to 25\% (14\%) in W-band (Q-band) by the uncertainty in the polarized flux from Tau A, measured most accurately by WMAP \citep{wmappol}. The Tau A measurements were noisier than the chopper plate results and were made at different times, but they did confirm the long timescale features discussed in \S\ref{sec:gains}.

\begin{figure*}
\centering
\includegraphics[width=\textwidth]{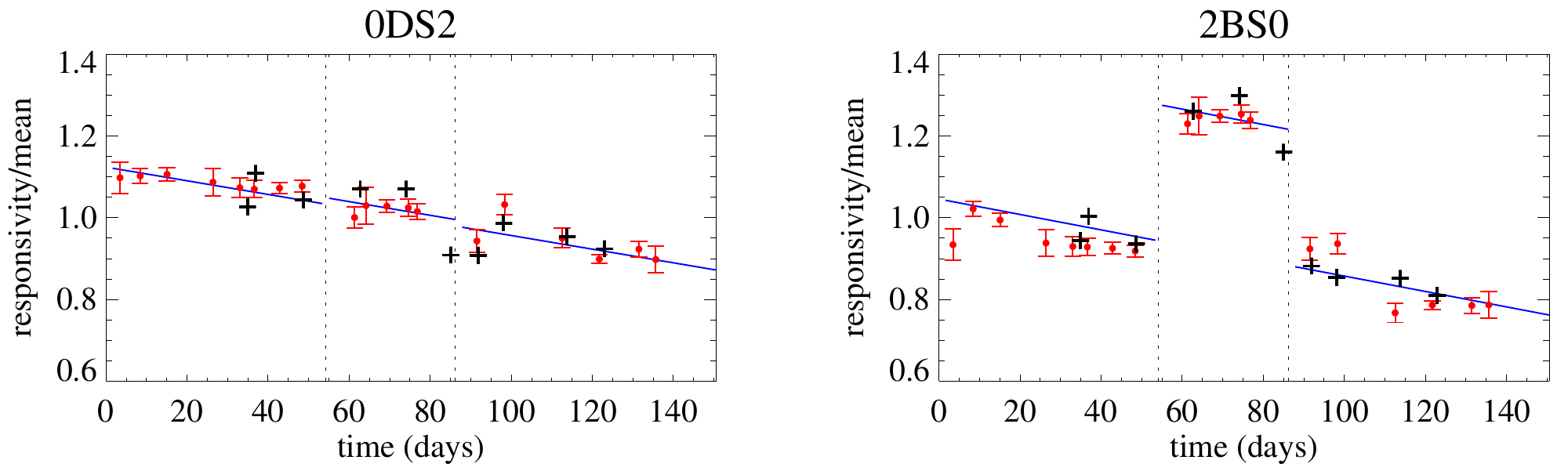}
\caption{\label{fig:gains} Responsivity variations during the observing season. Measurements for two polarization channels from the 10 chopper plate observations are plotted as plus symbols, with the $<1\%$ statistical error bars omitted. As discussed in the text, $\approx30\%$ of the channels exhibited a significant responsivity increase during period II, which is clearly visible for the channel on the right. The straight lines show the three-parameter baseline model (model 2 described in \S\ref{sec:gains}), which was used to interpolate the responsivity over the full season. The circles show the alternative model 1, in which the variations (other than the period II discrepancy) were predicted from measured temperature dependences. The model 1 points have been averaged into large bins during contiguous CMB observation runs, with the error bars indicating the spread in the predicted responsivity due to temperature variations during each run.}
\end{figure*}

As described in \S\ref{sec:beams}, the beam from radiometer 2Q was slightly truncated by the receiver cabin window, and so did not illuminate the chopper plate uniformly, which led to overestimation of the 2Q responsivities (since $Y$ in equation~\eqref{eq:chopP} was decreased by an unknown amount).  In fact, the 2Q responsivities inferred from the chopper plate measurements implied a much better sensitivity than measurements of the LNA noise  temperatures  allowed.  Thus, the responsivities for 2QS0 and 2QS1 were adjusted to match the Tau A measurements.

Since the temperatures of the active polarimeter components were known to have varied throughout the observing season, additional chopper plate observations were undertaken at the end of the season to measure the linear variation of the responsivity of each polarimetry channel to each of the three germane servoed temperatures described in \S\ref{sec:instrument}.  In each case, the responsivity decreased with increasing temperature.  The temperature coefficients varied for different channels but were typically 1~\%/K for the LNA and IF amplifier temperatures and 0.5~\%/K for the LO temperatures. 

\subsection{Responsivity Model}\label{sec:gains}
The chopper plate measurements revealed that the detector responsivities were not constant with time, with typical channels varying by  $20\%$ relative to their initial values, and the worst channels varying by as much as $60\%$.  We describe here the time-dependent detector responsivity model used for calibration of the CAPMAP timestreams.
 
Figure~\ref{fig:gains} shows the 10 chopper plate measurements for two representative channels. Since the temperatures of the active polarimeter components were seen to increase throughout the season, the  approximately steady decrease of the responsivities with time was not unexpected. In fact, considering the effects of all the temperature coefficients led to a typical prediction of a 20\% change across the full season, in good agreement with observations for most ($\approx70\%$) channels.  However, complete recovery of the responsivity variations on shorter timescales was prevented both by occasional periods of grounding problems in the readout thermometry, and by phase lags due to the poor placement of some thermometers relative to the active components.

The panel on the right in Figure~\ref{fig:gains} demonstrates an anomalous shift in the measured responsivity during period II. Such shifts affected about a quarter of the channels at the 10\% level or greater and were not forecast. We suspect the shifts, for which the worst case was 80\%, were related to the power outages at the boundaries of period II.  The sensitivity of each affected radiometer in mK$\sqrt{\mbox{s}}$ did not vary measurably across the periods---the shift was only a change in the V~K$^{-1}$ responsivity.  Finally, one radiometer, 1D, showed a season-long responsivity decrease in all three of its subbands three times larger than predicted from its temperature coefficients measured at the end of the season. The radiometer otherwise performed normally; this decrease was also only a change in the V~K$^{-1}$ responsivity.

To properly track the time dependence of the responsivities of the channels throughout the run (and to gauge our uncertainties in the understanding of said time dependence), we evaluated two different models for fitting the 10 chopper plate measurements and interpolating to each 21~s ring cycle.

For each channel, model 1's fit used the three measured temperature coefficients ($B_i$) and known housekeeping temperatures ($T_i(t)$) to constrain two parameters---$A$, an overall constant, and $A_{II}$, a possible offset for period II only:
\begin{equation}\label{eq:model1}
R_1(t) = A + A_{II} + \sum_{i=1}^3 B_iT_i(t).
\end{equation}

Model 2, now with three parameters per channel, replaced the last term in equation~\eqref{eq:model1} with a simple linear term: 
\begin{equation}
R_2(t) = C + C_{II} + Dt,
\end{equation}
where $D$ is a single negative slope (again different for each channel), characterizing the overall responsivity decrease with time, and $C$ and $C_{II}$ are constants analogous to $A$ and $A_{II}$ in model 1.

Both models were similar in that they accounted for the period II anomaly and for the gradual decrease of the responsivity throughout the observing season. Because model 1 applied the temperature coefficients to each ring cycle, it also addressed shorter-timescale changes in the responsivities.  However, as mentioned above, use of the temperature coefficients did not faithfully reproduce those changes in all instances.  For this reason, we used the simplified model 2, illustrated for two channels in Figure~\ref{fig:gains}, as our baseline model.  The model accounted  for the main features of the responsivity variations, with an RMS of 6.5\% for the 440 residuals from 44 channels and 10 chopper plate measurements. The systematic implications of its uncertainties are discussed in \S\ref{sec:systematics}.

\section{Data Analysis}\label{sec:analysis}
This section describes the steps in the data analysis culminating in polarization power spectra.  
We discuss first how the data were selected for processing, then their reduction into maps, and the subsequent estimation of power spectra. Two separate analysis pipelines (arbitrarily named 1 and 2) were developed;  at each step we discuss the approach of each. Throughout the analysis, we relied on simulations and consistency  checks to evaluate data quality, immunity to systematic effects, and internal consistency of the pipelines.  We conclude the section with descriptions of the simulation tools and the null test methodology.

\subsection{Data Selection} \label{sec:data_selection}
Since CAPMAP collected data from near sea level in the Garden State, it was critical to cull data contaminated by  large spatial and temporal fluctuations in the atmosphere during poor weather.
In this subsection, we describe all cuts made on the data and conclude with some observations about properties of the final data set.

We list here several preliminary cuts, indicating parenthetically the fraction of the 1658 hours of CMB data discarded by each. Data selection began by dividing the raw timestreams from each channel
into 21 s ring scan cycles. Only complete cycles were selected for inclusion in the final data set.
Cycles exhibiting irregular telescope motion were discarded (3.8\%), as were those during periods when the cryostats were warm or thermally unstable (6.9\%), or when known electrical or receiver malfunctions existed (2.4\%).  Only for the Q-band receivers, the sun passing through narrow, constant far sidelobes at specific times appeared as a several-mK signal in the polarimeters; cycles were cut according to the solar position (7.1\% of the Q-band data only). 

The remaining selection criteria were all derived from the receiver data, and were initially developed independently for the two pipelines.  Each used different estimators of the data quality.  In cutting, pipeline 1 rejected all polarimeters from individual cycles, while pipeline 2 made cuts on contiguous sets of 11 cycles and allowed for channel-specific cuts.  In the end, both pipelines converged to a common technique described below: pipeline 1's quality estimator applied on both timescales, with channel-specific cuts where appropriate. 
In distinguishing among differing techniques, we were guided by the results from the null test suite described in \S\ref{sec:null_tests} and \S\ref{sec:results}, not by the power spectra of the CMB maps.

To further characterize each cycle, for each total power (TP) or polarimetry channel we averaged the data into $N_\theta$ angular bins and fitted to the following five-parameter model:
\begin{equation}\label{eqn:modes}
M(\theta) = A_0 + A_1\sin\theta + A_2\cos\theta + A_3\sin 2\theta + A_4\cos 2\theta.
\end{equation}
For data selection, we used $N_\theta=18$. (In subsequent analysis steps, the data were binned into 200 or more angular bins, depending on the pipeline.) 

The first two terms in this Fourier mode expansion dominated the response of every channel: $A_0$ describes the TP system temperature (or the polarimeter offset), and $A_1$ the channel's first-order response to elevation changes, and thus to $T_{\text{atm}}$ as in equation~\eqref{eq:opacity}.
Their values for TP  channels were typically 100~K and 500~mK respectively;  for the polarization channels they were $\le150$~mK and $\le 5$~mK, governed by the size of the $I\rightarrow Q$ leakage.  For a single cycle, the remaining three terms were typically consistent with zero. Of these, 
$A_2$ and $A_4$ characterize an azimuthal dependence, while $A_3$ is the second-order term in the elevation dependence.

All subsequent cuts were  based on these fits for each cycle, most often on their  $\chi^2$ goodness-of-fit statistics. The primary cut variable used to reject bad weather was obtained from $\chi^2$ values from a subset of the TP channels, comprising the nine W-band channels with the lowest $1/f$ noise, constrained so that at least one such channel was in every cryostat.  (Since all TP channels had $1/f$ noise in excess of white noise, their mean $\chi^2$ values were all in excess of 13, the number of degrees of freedom.) For each subperiod, the nine $\chi^2$  distributions were renormalized to have their peaks at the same values, and  then the renormalized $\chi^2$ statistics were averaged to provide the cut variable for each cycle, a variable we call the TP-$\chi^2$.  Atmospheric contamination in the Q- and W-band TP channels was sufficiently correlated that the TP-$\chi^2$ also worked well for Q-band selection.

\begin{figure}
\centering
\includegraphics[width=3.39375in]{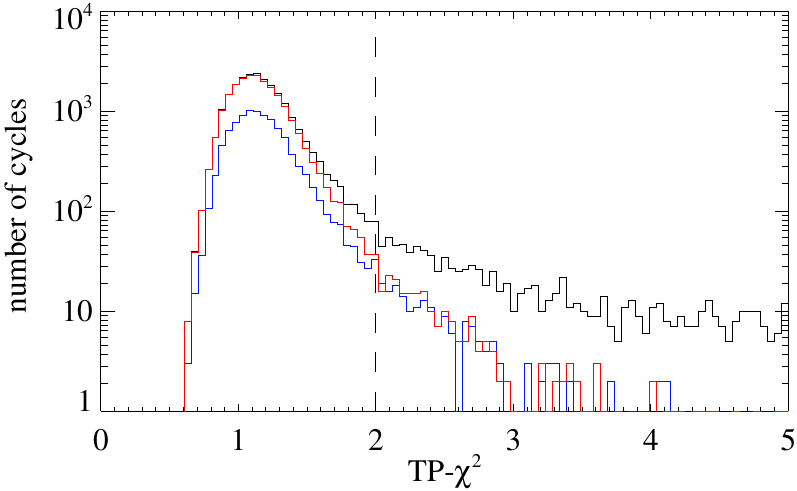}
\caption{\label{fig:tp_weathercut}Histogram of the cut variable TP-$\chi^2$ defined in the text. The black histogram shows the entire RP (24,000 contiguous 21~s cycles). A 10,000-cycle \emph{best-weather} subset (blue histogram) was identified by eye from the timestream shown in Fig.~\ref{fig:tp_weathercutstat}.
The distribution drawn in red was obtained from the entire RP distribution after applying just the 11-cycle cut (see text); it approximates the best-weather distribution. The single-cycle cut (vertical dashed line) removed the long tail.}
\end{figure}

\begin{figure}
\centering
\includegraphics[width=3.39375in]{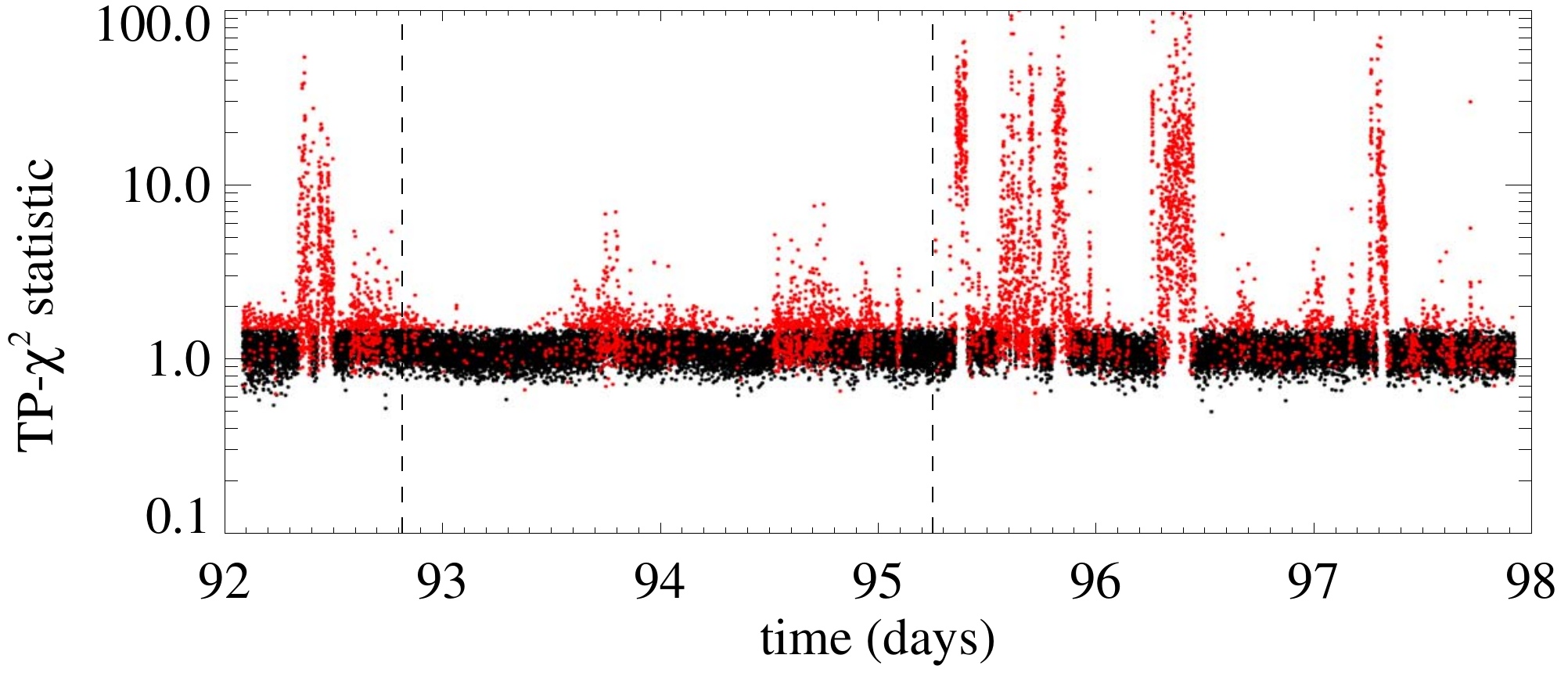}
\caption{\label{fig:tp_weathercutstat}Timestream of the TP-$\chi^2$ cut variable defined in the text, during the RP.  Cycles eliminated by the two-timescale weather cut described in the text are indicated in red. The RP was chosen to be a particularly quiet period with no gaps, so a smaller fraction of cycles was cut than was typical. The dashed lines mark the best-weather subset.}
\end{figure}

The RP (see \S\ref{sec:cmbobs}) was a particularly clean six-day period of data identified early in the analysis and studied extensively for preliminary work developing the TP-$\chi^2$ as a cutting variable. For reference, the RP is indicated in Figure~\ref{fig:tau}. Figure~\ref{fig:tp_weathercut} plots the distribution of the TP-$\chi^2$ during the RP, while Figure~\ref{fig:tp_weathercutstat} plots its timestream and gives an indication of its power for data selection.

\begin{figure*}
\centering
\includegraphics[width=\textwidth]{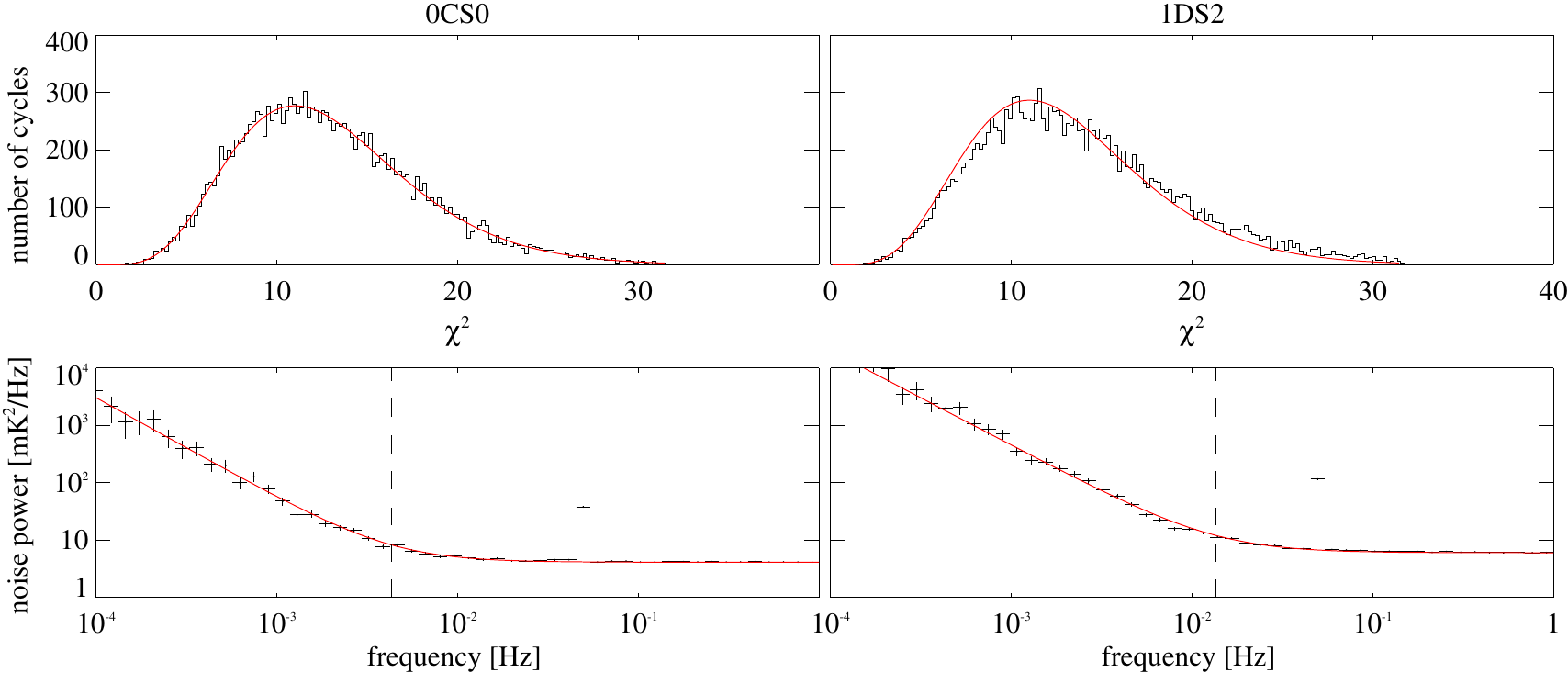}
\caption{\label{fig:5m_csq}The distribution of the five-parameter $\chi^2$ statistic is shown in the upper panels for cycles from one of the nine subperiods, with about 80 hours of data, for two polarimeter channels after selection cuts. A $\chi^2$ distribution with 13 degrees of freedom is overlaid as a smooth curve. The channel on the left, a typical one, has a mean $\chi^2$ of $12.95$, consistent with the expected value $12.94\pm0.04$ (for a distribution truncated at $3\sigma$), while the one on the right, one of the worst channels, has a mean $\chi^2$ of $13.62$, which is discrepant at more than $16\sigma$.
The lower panels show the noise power spectra for the same two channels. The solid line is a three-parameter fit: the white noise level, the knee frequency, and the slope of the $1/f$ component. The dashed vertical lines indicate the knee frequency of each channel (4.3 and 13.6 mHz for the left and right panels, respectively); the ring scan frequency can be clearly seen as a high point at roughly 50 mHz. Higher levels of $1/f$ noise in the channel on the right increase the variance at the ring scan frequency by 12\% over the variance of 100 Hz samples, inflating the $\chi^2$ statistic. Of 44 data channels, only 10 had knee frequencies greater than 10 mHz. Although the five-parameter $\chi^2$ statistic is extraordinarily sensitive to $1/f$ noise, systematic studies showed the effect of $1/f$ noise on the full-season results to be negligible (\S\ref{sec:systematics}).}
\end{figure*}

As indicated previously, selection on two timescales sharpened the effectiveness of the cuts. The TP-$\chi^2$ distribution in conjunction with results from the null test suite (\S\ref{sec:results}) was studied with many combinations of parameters defining the single-cycle cut and the longer timescale cuts. The chosen parameter set was: any 11 cycles were removed if at least seven of them had a TP-$\chi^2$ $>1.5$; and any single cycle was cut if its TP-$\chi^2$ exceeded 2. These cuts were applied uniformly to all polarimetry channels, discarding 19.4\% of the full data set.
 
The five-ring-mode fits to the polarization channels were also used for the final cuts on the data.\footnote{We used simulations to verify that the presence of the CMB signal in the timestreams did not in any way bias the selection of cycles.}  For any candidate cycle, the sum of the $\chi^2$ values  for all functioning W-band (and, separately, Q-band) polarimeters was formed, and if either of these sums was  $\geq3\sigma$ high, that cycle was cut for {\it all} polarimeters. Similar cuts were made on the maximum residual to the five-parameter fit for all functioning W-band (and separately Q-band) polarimeters in a given cycle.  For these two cuts,  the idea was to use the overall goodness of fit for the ensemble of polarimeters in a given cycle  to keep either all or none of them.  A further 1.6\% of the data was so removed.  The next cut was based upon the fact that the noise of a properly functioning polarimetry channel should depend linearly on the total power level.  This relationship was examined for each polarimeter, for each of the nine subperiods.  The linear relationships were dominant, but for some isolated periods, cycles were found to deviate significantly from the expected behavior.  As these deviations indicated receiver malfunctioning, those cycles were discarded (3.5\%). A final cut was invoked to remove glitches by eliminating cycles for individual polarimeters with a residual deviant at more than 3$\sigma$, removing 1.7\% of the Q-band and 0.3\% of the W-band data.

We close this subsection with some observations about the data quality of the polarimeters derived from examination of their five-ring-mode fits.  In particular, we define two measures that were not used in data selection but were used to classify data in the null test suite (\S\ref{sec:results}). 

In contrast to the TP channels, the $\chi^2$ distributions from the fits to most polarimeter channels did closely follow the expected $\chi^2$ distribution for 13 degrees of freedom, verifying at the same time our noise model (white noise) and that five parameters described the data well. One such channel's distribution is shown in the top panel on the left in Figure~\ref{fig:5m_csq}. A few channels, however---including the one depicted on the right in Figure~\ref{fig:5m_csq}---exhibited $\chi^2$ distributions shifted slightly high. Given the large number of cycles entering these distributions, even a $\chi^2$ shift of less than 1.0 was significant at more than 10 standard deviations. It was found, not surprisingly, that such channels  were the ones that showed the worst (highest)  $1/f$ knees;  and that such channels were the same in every one of the 9 subperiods. For comparison, Figure~\ref{fig:5m_csq} shows a noise spectrum for each of the two channels in the lower panels.  We defined a polarimeter's \emph{grand $\chi^2$} as the mean value of its $\chi^2$ distribution; we tested for contamination associated with high grand $\chi^2$ in \S\ref{sec:results}. We also simulated the effects of $1/f$ noise directly, as described in \S\ref{sec:systematics}.

For each polarimeter channel, and again for each subperiod, we constructed the 18-element \emph{ground-synchronous structure} vector from an appropriate average of the residuals in each $\theta$ bin of the ring scan. We evaluated the $\chi^2$ for the model of zero ground-synchronous structure for each polarimeter; the W-band channels were consistent with no such signal, but two of the eight Q-band channels were not. Those two channels showed no evidence, however, for time-variability in their ground-synchronous structure. Scan-synchronous pickup from telescope sidelobes would be manifest as nonzero ground-synchronous structure, so these results are further evidence that the telescope sidelobe reduction efforts (\S\ref{sec:instrument}) were largely successful. Nonetheless, as described in \S\ref{sec:mapmaking}, the ground-synchronous structure mode was projected out of the data before power spectrum analysis. 

\subsection{Mapmaking}\label{sec:mapmaking}
The timestream of a given CAPMAP channel can be decomposed into signal plus noise as follows:
\begin{equation}
\vect{d} = P\vect{s} + \vect{n}.   \label{eq:tod_sn}
\end{equation}
Here, \vect{d}\ denotes the timestream values, represented as a vector of length $\Ntod$.
The celestial signal in the map space is represented as a vector \vect{s}\ with $2\Npix$ components $s_p$, where
the index $p$ denotes a combination of a pixel and a choice of Stokes parameter.
The $\Ntod\times2\Npix$ pointing matrix is denoted $P$. The coordinate system in which \vect{s}\ is defined and the corresponding form of the pointing matrix were different for the two pipelines, as discussed below.

The objective of mapmaking is to invert equation~\eqref{eq:tod_sn} statistically, arriving at an
estimator $\mathbf{x}$ for the CMB signal $\mathbf{s}$, with associated noise covariance matrix $N$. The estimated signal $\mathbf{x}$ is called a \emph{map vector}. 
It is well known that there is an optimal choice of unbiased estimator (e.g., \cite{Tegmark:1996qt}), 
defined by the pair of equations
\begin{equation}  \label{eq:capmap_npq_optimal}
\begin{split}
N^{-1} \mathbf{x} &= P^\dagger C^{-1} \vect{d} \\
N^{-1}  &= P^\dagger C^{-1} P \,,   
\end{split}
\end{equation}
assuming that the detector noise \vect{n}\ can be modeled as Gaussian with covariance
matrix $C = \bigl\langle \vect{n}\vect{n}^\dagger\bigr\rangle$.

The covariance matrix $C$ needed in equation~\eqref{eq:capmap_npq_optimal}
was derived from a noise model that assumed white noise. Then, certain modes in the data that were to be marginalized in the mapmaking were assigned large variance as follows:
\begin{equation}
C = C_0 + \epsilon^{-1} R^\dagger R + \epsilon^{-1} G^\dagger G,   \label{eq:capmap_ns_again}
\end{equation}
where $C_0$ was diagonal, $\epsilon$ was a small regulating constant,
and each row of the matrices $R,G$ contained one timestream mode to be projected out. The assumption that $C_0$ was diagonal---that is, that the receiver noise was uncorrelated at different times---was verified by data quality studies.
The white noise level used in $C_0$ was independently estimated in each cycle to allow for time dependence. The noise estimation method was tested extensively in simulations and found to reconstruct the true white noise level with acceptable accuracy. The matrix $R$ contained ring modes: the lowest five Fourier modes in the ring variable $\theta$ in each cycle. The matrix $G$ contained ground-synchronous structure modes: one mode for each
of $N_\theta$ bins around the ring scan.

Assigning a large variance to the ring and ground-synchronous modes was merely a device for
projecting them out when making $N^{-1}$-weighted maps (eq.~\eqref{eq:capmap_npq_optimal}).
Those modes would be removed completely in the limit $\epsilon\rightarrow 0$. It was preferable to keep the noise matrix invertible, however, by assigning a small nonzero value to $\epsilon$,  chosen to be large enough to avoid numerical difficulties but still insignificant compared to typical elements of $C$. We verified that changing this parameter over three orders of magnitude did not affect the results of either pipeline.

Ring mode removal was necessary to project out the contaminating atmosphere signal from $I\rightarrow Q$ leakage, but it also served to highpass filter the timestream, so that the remaining noise could be treated as white. The mode removal led to an increase of the statistical error on the final result in the full range $2\leq\ell\leq3000$ by 15\%. Most $\ell$ bands were affected only by 5--10\%, but the sensitivity of the lowest band, $2\leq\ell\leq500$, was reduced by 70\%, and the sensitivity for $\ell\leq200$ was essentially eliminated.

The noise was verified to be uncorrelated between different channels, including different subbands of each polarimeter, and consequently we estimated a net data vector \vect{x}\ for a given subset of W- or Q-band channels by summing the individual noise-weighted maps (eq.~\eqref{eq:capmap_npq_optimal}) and adding the inverse noise covariance matrices. We also combined W- and Q-band channels in an enlarged pixel space (\emph{Q$+$W analysis}) as described in \S\ref{sec:ps}.

In pipeline 1, the pixel space was a standard pixelization in declination and right ascension (RA), using HEALPix\footnote{\url{http://healpix.jpl.nasa.gov}} with $N_{\text{side}}=2048$. At this resolution there were $\approx10^4$ pixels in our field, each of diameter $1.7\arcmin$ or roughly one-half the W-band beam FWHM. Mapmaking by brute force application of equation~\eqref{eq:capmap_npq_optimal} would require
$\bigoh(\Ntod^3)$ matrix operations and therefore be prohibitively expensive.
However, we found a mathematically equivalent algorithm \citep{Smith:07T}, based on repeated application of the Sherman-Woodbury formula \citep{sherman}, that was computationally efficient.

In pipeline 2, the underlying pixel space was an azimuthal pixelization about the NCP as described by \citet{crittenden}, with $N_\delta=90$ bins in declination ($\delta$) and $N_\phi=512$ bins in RA.  Then, the largest pixels were at the outer edge of the survey region, with diameter $1.2\arcmin$ or one-third of the W-band beam. 
However, estimation of $\mathbf{x}$  was obtained in  the half-Fourier space ($\delta$, $k_\text{RA}$), where here and below the symbol $k_v$ represents  the Fourier transform variable for a coordinate $v$.  
Prior to construction of the pointing matrix $P$, the timestream data from each polarimetry channel were binned into  ($\theta$, LST) coordinates.  Data at fixed $\theta$ were Fourier transformed in LST, resulting in a  \emph{receiver map} in ($\theta, k_\text{LST}$), used as $\vect{d}$ in equation~\eqref{eq:capmap_npq_optimal}.
Each $\theta$ corresponded to a fixed declination, hour angle, and parallactic angle, so $k_\text{RA}$ only differed from $k_\text{LST}$ by a fixed phase for each $\theta$, and $P$ was straightforward to compute. Ring mode removal was performed as in pipeline 1, using equation~\eqref{eq:capmap_ns_again}, though without the
term $G^\dagger G$. Ground-synchronous structure corresponds to an LST-independent signal, which was removed in these coordinates by discarding the DC mode $k_{\text{RA}}=0$.

The use of the Fourier transform greatly accelerated the mapmaking process, with evaluation scaling as $\bigoh(N_\phi N_\delta^3)$ rather than $\bigoh(N_\phi^3N_\delta^3)$.  This acceleration came about because in this basis the noise covariance matrix was block diagonal---it did not connect pixels with different values of $k_{\text{RA}}$---in the limit 
that the noise in each channel and each $\theta$ bin was independent of LST. In fact the noise was only LST-invariant at the $\approx10\%$ level, but extensive simulations showed that
the approximation of invariant noise resulted in a negligible systematic error,
as discussed in \S\ref{sec:systematics}. The half-Fourier coordinate system also diagonalized the theory covariance matrix, yielding a similar speed increase for the power spectrum analysis. The power spectrum estimation thus broke  down into $N_\phi/2$ separate subproblems, one for each unique value of $k_{\text{RA}}$. Further acceleration by a factor of three was obtained by discarding Fourier modes with $k_{\text{RA}}>80$, which only probe angular scales beyond $\ell=3000$.

\subsection{Power Spectrum Analysis}\label{sec:ps}
The power spectrum analysis was performed by calculating the likelihood \like,
\begin{equation}\label{eq:like}
-2\ln\mathcal{L} = \ln \det (N+S) + \mathbf{x}^\dagger (N+S)^{-1}\mathbf{x},
\end{equation}
with respect to the theory encoded by $S$. The theory covariance matrix $S$ can be written
\begin{equation}
S = \sum_{i=1}^M \alpha_i\xi_i,
\end{equation}
where $\alpha_i$ represent $M$ parameters to be estimated from the data and the theory basis matrices $\xi_i$ are calculated from  trial $EE$ and $BB$ (and sometimes $EB$) power spectra
using standard formulas \citep{zaldarriaga}. Results for the parameters $\alpha_i$ are reported in \S\ref{sec:results} for several different choices of the number of bands $M$ and the trial power spectra.

To estimate the overall sensitivity and to evaluate the agreement of the data with the concordance model for $EE$ defined in \citet{WMAP3} and evaluated using CAMB,\footnote{\url{http://camb.info/}} 
we performed two-band fits with one $EE$ and one $BB$ band in the range $\ell\in[2,3000]$.
For these, the $EE$ parameter was a dimensionless multiplier to the concordance model,
while the $BB$ parameter was a flat bandpower in $\mu$K$^2$.
For evaluating null maps (see \S\ref{sec:null_tests} and \S\ref{sec:results}), we made similar two-band fits, but with both the $EE$ and $BB$ parameters taken as flat bandpowers.

To estimate spectral features for the final analysis, we formed seven bins in $\ell$ with width $\Delta\ell=300$, except we expanded the lowest bin ($2\leq\ell\leq500$) to account for mode removal, and we extended the highest bin out to $\ell=3000$ to provide a check for unexpected power (perhaps from point sources) on small angular scales.  The choice $\Delta\ell=300$ resulted in small correlations between adjacent bandpowers (typically 5\%, and 20\% for the worst case);  the correlations rose sharply for narrower bins because of mode-coupling from the finite size of the survey region.  We estimated $EE$ and $BB$ for each bin as flat bandpowers, resulting in 14 bands in total.

For the Q$+$W analysis, the pixel space was expanded by a factor of two, so that each
map pixel appeared once for each of the two different beam sizes. The noise was taken, as usual, to be uncorrelated 
among all polarimetry channels, 
but the theory matrix was constrained to represent the same underlying CMB 
sky at both frequencies (properly accounting for the beam window function for cross-correlating two beams with Gaussian widths $\sigma_1,\sigma_2$: $B_\ell(\sigma_1,\sigma_2) = \exp\bigl[-\ell(\ell+1)(\sigma_1^2+\sigma_2^2)/2\bigr]$).

From the $M$-dimensional joint likelihood function $\like(\alpha_1,\ldots,\alpha_M)$, we calculated the $M$ one-dimensional marginalized functions $\like_i(\alpha_i)$, from which all statistical inferences followed. Each result is quoted as the maximum likelihood value with an asymmetric $1\sigma$ confidence interval calculated as the 68\% interval of highest density. For $BB$ bands we also calculated 95\% confidence upper limits; the limits were derived using only the portion of the likelihood curves in the physical region of positive bandpower.

The calculation of the joint likelihood was accelerated by the method of signal-to-noise eigenmodes \citep{Bond:1998}, which transforms equation~\eqref{eq:like} into a simple algebraic expression that determines the likelihood on any dense one-dimensional subset of the $M$-dimensional parameter space with only two expensive matrix operations. For the two-band analysis, we calculated the joint likelihood on a 60-element grid of such subspaces and obtained the marginalized likelihoods by numerical integration. For larger $M$, a useful procedure was to use the signal-to-noise eigenmode method to obtain $M$ \emph{conditional likelihoods} $\like(\alpha_i|\alpha_{j\neq i})$, in which the likelihood for the $i$th band was calculated as a one-dimensional function with the other $M-1$ bands fixed to the maximum likelihood location. These were only approximations to the true marginal distributions, but in practice they were not much different and could be computed with significantly less effort; they were valuable both as diagnostic tools and as components of more complicated analyses. We found that the marginalized likelihoods in the two-band $EE/BB$ case were not significantly different from the conditional likelihoods, which was expected from the orthogonality of $EE$ and $BB$ due to the uniform parallactic angle coverage of the scan strategy. The correlation between the $EE$ and $BB$ bands was less than $5\%$.

We assigned errors to $EB$ bandpowers by fixing $EE$ and $BB$ to their maximum likelihood values (computed assuming $EB=0$) and then computing conditional likelihoods in $EB$, rather than marginalizing over $EE$ and $BB$.
The reason is that we found $EB$ bandpowers to be correlated significantly with the others, but there was a strong prior belief that $EB=0$. Likewise, when we assigned errors to $EE$ and $BB$ bandpowers, 
we did not include additional uncertainty from marginalizing over $EB$.

For the 14-band analysis, we sampled the marginal distribution using Markov Chain Monte Carlo (MCMC). The specifics differed for the two pipelines and are discussed briefly in the following; further details are given in \citet{Smith:07T} and \citet{lewis_thesis}.

\subsubsection{Pipeline 1}
Since the signal and noise covariance matrices were dense matrices of size $n\approx (2\times 10^4)$, and likelihood operations typically involve $\bigoh(n^3)$ matrix operations, several optimizations were required to make the MCMC analysis computationally feasible.

The first optimization was the use of signal-to-noise eigenmodes to reduce the effective matrix size; after a few $\bigoh(n^3)$ operations to 
pre-compute the eigenmodes and signal covariance matrices, subsequent likelihood operations could be performed using matrices of
smaller size $n_{\text{ev}} \ll n$. For CAPMAP, we found that no information was lost by choosing $n_{\text{ev}}=4500$, reducing the dimensionality by more than a factor of $4$.

The efficient calculation of conditional likelihoods enabled a fast MCMC technique using Gibbs sampling. The joint likelihood $\like(\alpha_1, ..., \alpha_M)$ was sampled
by sampling the conditional likelihoods repeatedly.  After the $n$th iteration, we had a new MCMC sample $(\alpha_1^{(n)},\cdots,\alpha_M^{(n)})$ from the joint likelihood, 
and we saved the conditional likelihoods $\like(\alpha_i|\alpha_{j\ne i}^{(n)})$.
One can show that the conditional likelihood averaged over MCMC samples, 
\begin{equation}
\frac{1}{N_{\rm samples}} \sum_{n=1}^{N_{\rm samples}} \like(\alpha_i|\alpha_{j\ne i}^{(n)}),
\end{equation}
converges to the marginalized likelihood $\like(\alpha_i)$.
By generating a full conditional likelihood in each iteration, rather than just generating samples, the rate of convergence was dramatically improved.
We found that only $\approx 10^3$ MCMC samples were needed to obtain convergence.

\subsubsection{Pipeline 2}
We used the standard Metropolis-Hastings algorithm \citep{metropolis,hastings} with a Gaussian proposal density. We matched the proposal density closely to the actual marginal distributions by first calculating the conditional likelihoods, obtaining acceptance probabilities as high as 70\%. We used 350000 MCMC samples to obtain the final results, although convergence was typically achieved with 3--5 times fewer samples.

Because of the Fourier space optimization, it was also feasible to calculate the joint likelihood by a brute force evaluation of equation~\eqref{eq:like} on a $60^M$-element grid for $M\leq4$. This enabled us to verify that our MCMC implementation correctly reconstructed the marginalized likelihoods.

\subsection{Simulations}
All aspects of both pipelines were tested extensively on simulated data sets. To create a \emph{full-season simulation}, we processed the complete data set, replacing each 100-Hz sample for all 44 channels and all 280000 cycles with a data point drawn at the same RA, declination and parallactic angle from a simulated CMB realization of the CAPMAP field, with Gaussian noise added at the same level present in the real data at that time. The simulated timestreams were then processed into simulated map vectors by the binning and mapmaking codes, applying the same data selection used for the real data. The resulting map vector had the same white noise properties as the actual data set, but a known underlying CMB signal, which could be set to have any desired power spectrum, typically the concordance model for $EE$ and zero power for $BB$.

These simulations enabled end-to-end tests of the binning, noise estimation, mapmaking, and power spectrum estimation schemes. The maximum likelihood estimator algorithms were verified to be unbiased at the subpercent level (much smaller than the statistical error bars), and the asymmetric confidence intervals were verified to have the expected statistical properties, in all $\ell$ bands. Timestream simulations were exchanged between the two pipelines, verifying that they were consistent with each other as well as internally.

For pipeline 2, it was also convenient to generate simulated receiver maps in the ($\theta$, LST) binning space directly for many studies.  Each such simulation could be generated in only a few minutes, and the mapmaking and power spectrum estimation  proceeded with similar speed in the half-Fourier space.

\subsection{Null Tests}
\label{sec:null_tests}
The pipelines were designed to allow a suite of  \emph{null tests}---statistics that explored a large set of possible sources of contamination and systematic uncertainty. The first step was to
make separate maps $(N_1,\mathbf{x}_1)$ and $(N_2,\mathbf{x}_2)$ for two disjoint subsets of the data (where $N_i$ denote the noise covariance matrices and $\mathbf{x}_i$ denote the map vectors). We then
constructed a null map $(N,\mathbf{x})$ as:
\begin{equation}\label{eq:nullprocedure}
\begin{split}
N &= N_1 + N_2   \\
\mathbf{x} &= \mathbf{x}_1 - \mathbf{x}_2,
\end{split}
\end{equation}
and we performed a two-band power spectrum analysis on the null map, fitting both $EE$ and $BB$ as flat bandpowers. The null maps were constructed to have no contribution from the CMB and thus be consistent with zero signal. This procedure tested for systematic contamination associated with the way in which the data were divided into subsets, e.g.\ sun contamination in the case of a day/night split.
Furthermore, the null maps could be examined at early stages of the analysis without biasing the final results.

The differencing procedure defined by equation~\eqref{eq:nullprocedure}
is appropriate when the two maps have the same spatial resolution.
This was the case for almost all of our null maps, but there was one important exception: the \emph{Q$-$W null map}, which tests for overall agreement between the two frequency bands. In this case, we handled the distinct beam sizes by degrading the W-band map to the coarser Q-band beam size before differencing using equation~\eqref{eq:nullprocedure}. The degrading operation was performed directly in map space and applied consistently to the noise covariance matrix as well as the map vector.
Since the null map was limited already by the coarser resolution, no information was lost in this procedure. Map differencing with distinct beam sizes was only implemented in pipeline 1, because this procedure mixes $k_\text{RA}$ Fourier modes.

We summarized likelihoods resulting from null maps with an \emph{equivalent $\chi^2$}, which was obtained by calculating the probability to exceed zero and expressing it as a corresponding $\chi^2$ with one degree of freedom.

\section{Power Spectra}\label{sec:results}
In this section we present the power spectrum estimates for $EE$ and $BB$.
Power spectrum analysis of the Q$-$W difference map and fits for $EB$ power are also presented along with results of the comprehensive null test suite. Our primary measurements are of the $EE$ and $BB$ power spectra in the range $200 \lesssim \ell\lesssim 3000$.  

\begin{figure*}
\centering
\includegraphics[width=\textwidth]{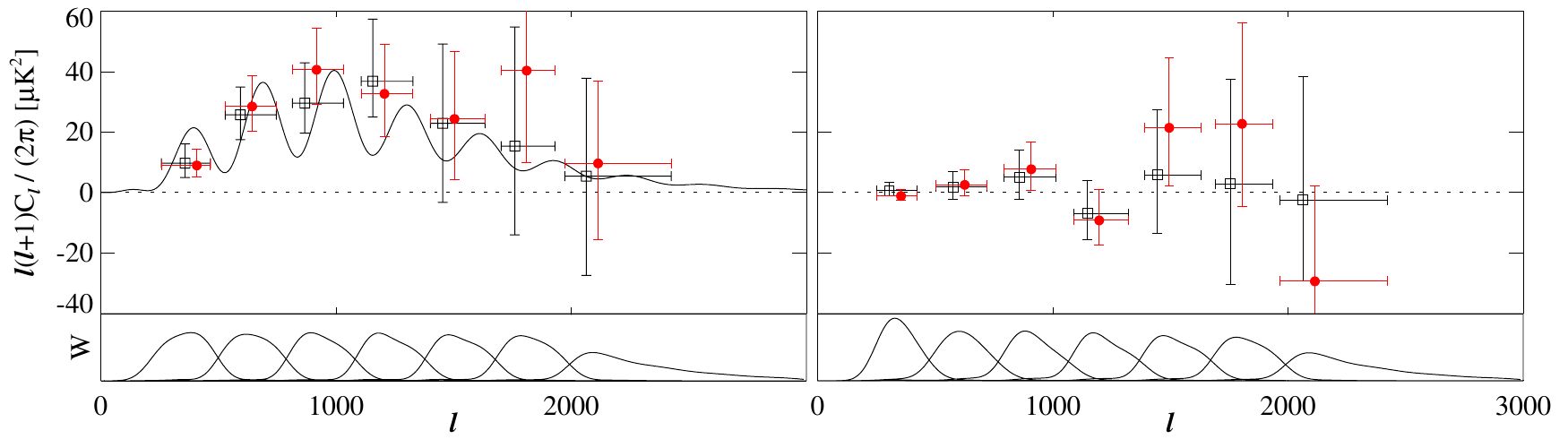}
\caption{\label{fig:results1}Q$+$W combined results from pipeline 1 (squares) and pipeline 2 (circles) for $EE$ (left) and $BB$ (right) flat bandpowers. The points for pipeline 1 (2) are shifted by $\ell=-10$ $(10)$ for clarity. The $EE$ concordance model is shown as the solid curve, and the window functions for each band are plotted below the power spectrum results. Neighboring bands have only small correlations (see \S\ref{sec:ps}). The error bars show the statistical errors only and do not reflect the $\approx20$\% overall calibration uncertainty or the small systematic errors discussed in \S\ref{sec:systematics}.}
\end{figure*}

\begin{figure}
\centering
\includegraphics[width=3.39375in]{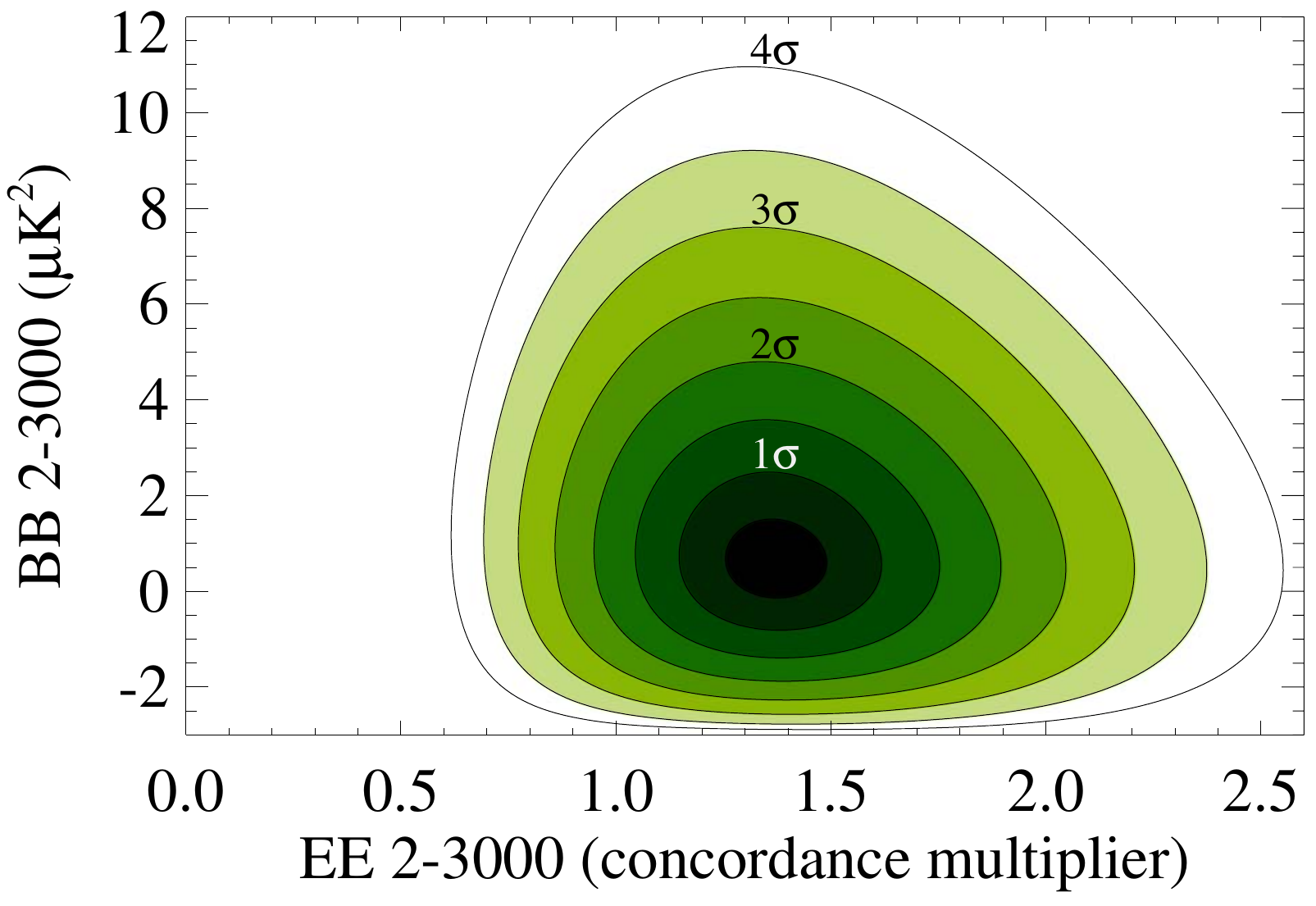}
\caption{\label{fig:2band}$EE$/$BB$ joint likelihood contours from pipeline 2 for the two-band Q$+$W fit. Contours are drawn at the one-dimensional $n\sigma$ levels: $\mathcal{L}_p\exp(-n^2/2)$, where $\mathcal{L}_p$ is the peak likelihood.}
\end{figure}

\begin{figure*}
\centering
\includegraphics[width=\textwidth]{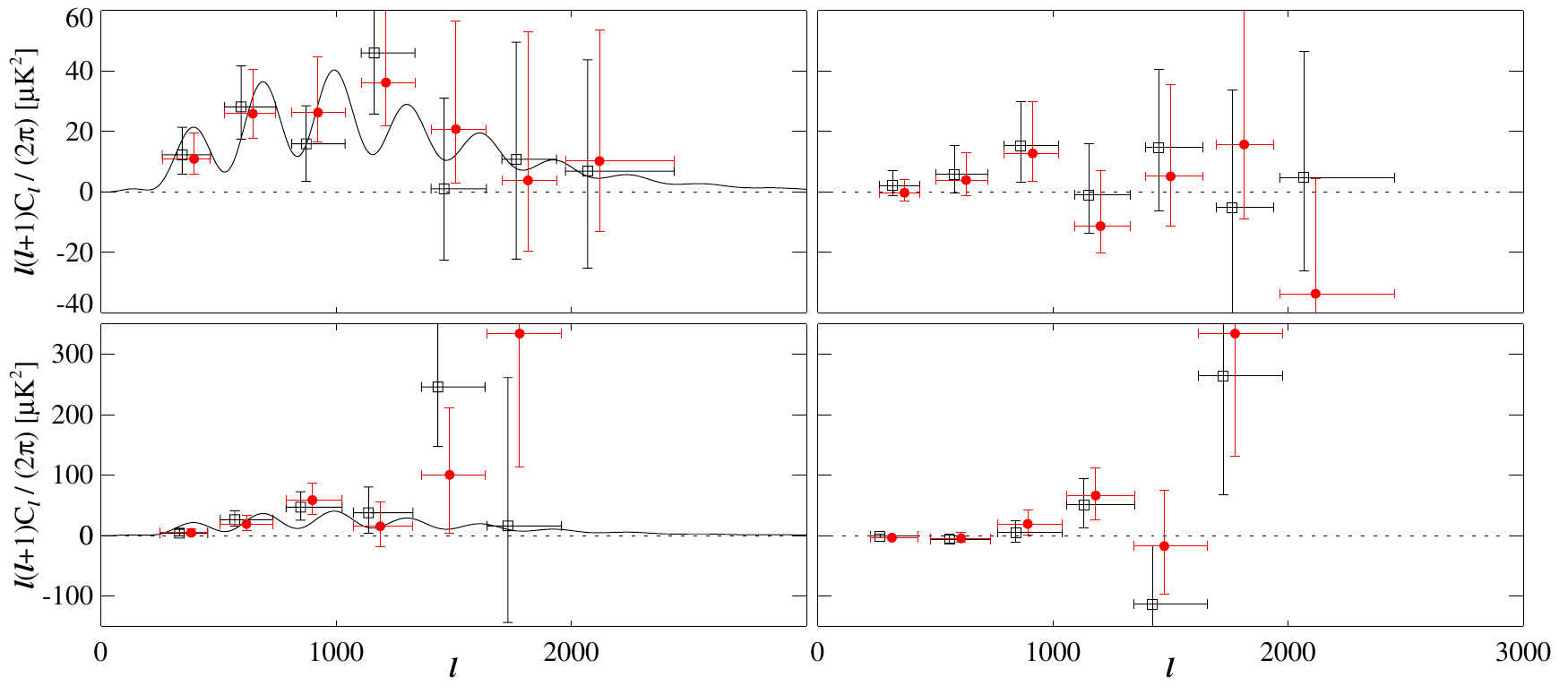}
\caption{W-band (top) and Q-band (bottom) power spectra from pipeline 1 (squares) and pipeline 2 (circles). $EE$ is on the left and $BB$ is on the right. We have omitted the $2000\leq\ell\leq 3000$ bandpower in Q-band, where the resolution is not adequate to estimate power on such small scales.}
\label{fig:results2}
\end{figure*}

\begin{figure*}
\centering
\includegraphics[width=\textwidth]{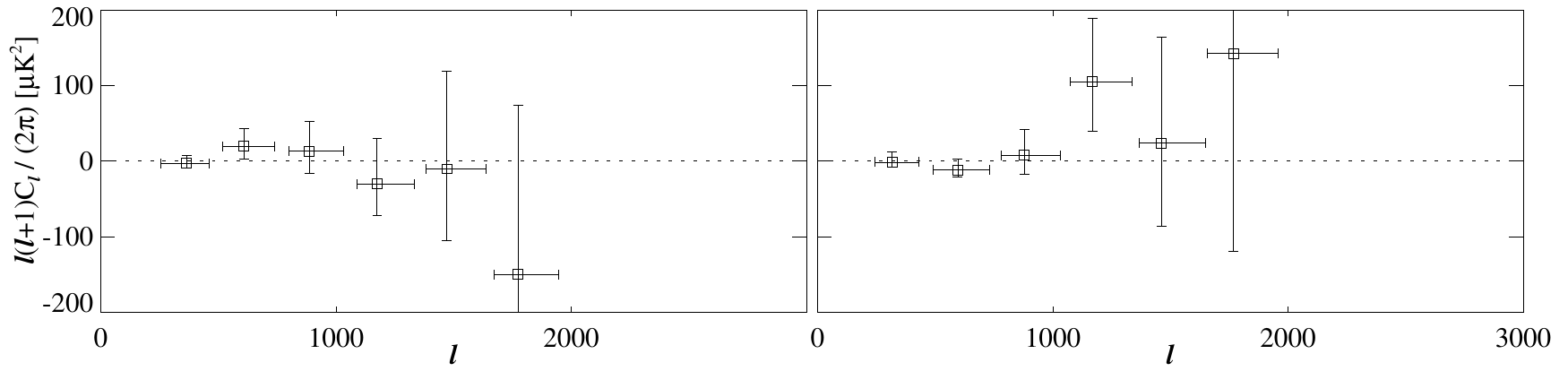}
\caption{Power spectra from pipeline 1 of the Q$-$W map. The lack of power in both $EE$ (left) and $BB$ (right) is a good indication that, within the noise level of this map, polarized foregrounds at either frequency are not contributing.}
\label{fig:results3}
\end{figure*}

\begin{figure}
\centering
\includegraphics[width=3.39375in]{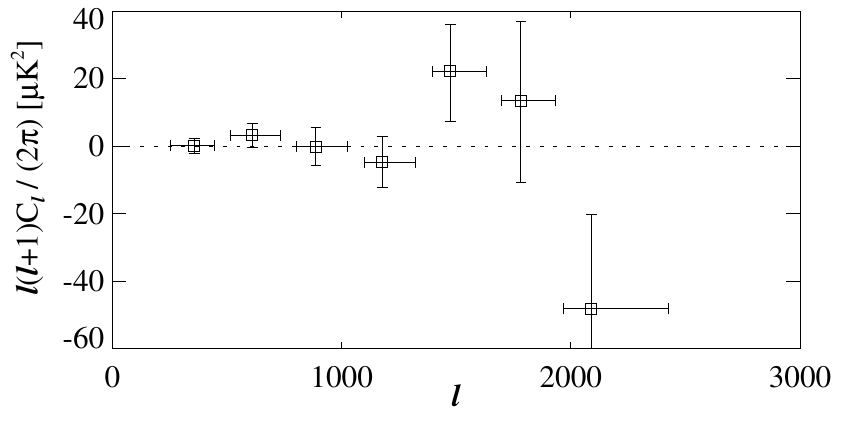}
\caption{\label{fig:EBresults}Estimated $EB$ bandpowers from pipeline 1 for the Q$+$W combined map, showing consistency with $EB=0$.}
\end{figure}

\begin{figure*}
\centering
\includegraphics[width=\textwidth]{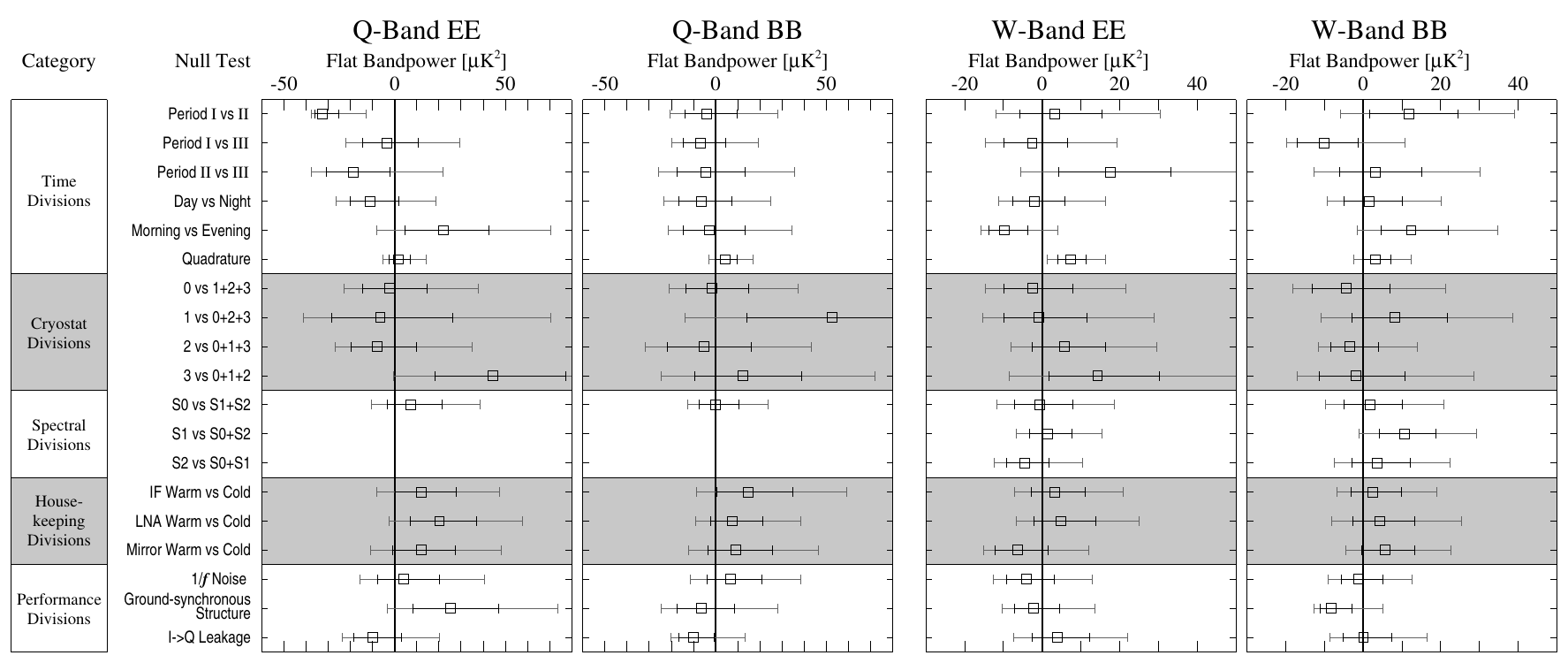}
\caption{\label{fig:nullsuite} Results from two-band fits to the suite of 34 null maps for Q-band (left) and 38 for W-band (right). Flat bandpowers with $1\sigma$ and $2\sigma$ error bars are shown for each. Each null test is described further in the text.}
\end{figure*}

\begin{figure}
\centering
\includegraphics[width=3.39375in]{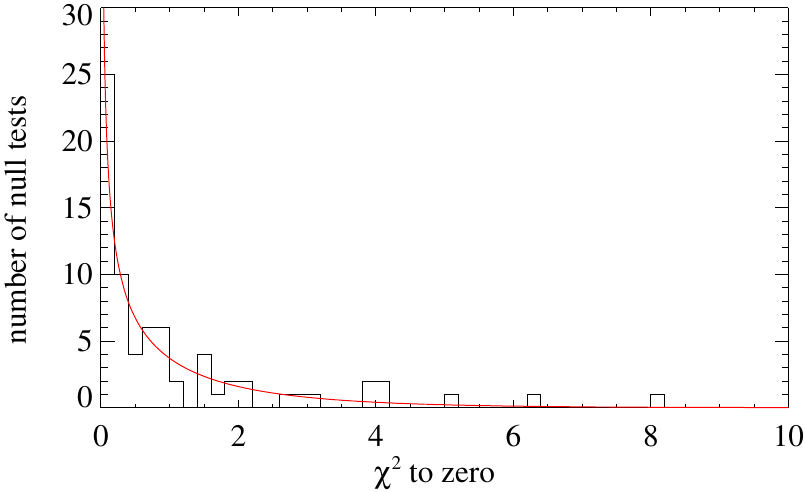}
\caption{\label{fig:nullchis}The $\chi^2$ distribution for fitting each test in the CAPMAP null test suite. The smooth curve shows the theoretical distribution for one degree of freedom. The agreement indicates that the ensemble of null tests shows no evidence for contamination.}
\end{figure}

\paragraph{Q$+$W Spectra} In Figure~\ref{fig:results1} we present results from each pipeline for a likelihood analysis of the Q- and W-band maps together. Although the two pipelines were independently developed and implemented, and differed in key details (including pixel size and shape, map space, and noise modeling) as described in \S\ref{sec:analysis}, the two sets of results agree very well. The small variations between the two are consistent with expected statistical variations from the different binning and mapmaking procedures, as verified by full-season simulations exchanged between the two pipelines.

The results also agree with the concordance model: the $EE$ power is significantly detected in multiple bands and $BB$ power is consistent with zero. The window functions for the seven bands have been calculated as in \citet{knox_window} and are plotted in panels under the power spectra in Figure~\ref{fig:results1}. The effective $\ell$ for each band is taken as the maximum of the window function, and the horizontal error bars indicate the 68\% highest-density confidence interval. At high $\ell$, almost all the statistical weight came from W-band because of its smaller beam size; at low $\ell$, the weight was slightly dominated by Q-band because of its larger survey area.

\paragraph{Two-band Spectra} Figure~\ref{fig:2band} shows the joint likelihood contours obtained by one of the pipelines
for the $EE$/$BB$ two-band analysis. The $EE$ band yielded an $11\sigma$ detection with a concordance multiplier $1.36^{+0.25}_{-0.22}$. For $BB$ we obtained a flat bandpower $0.6^{+1.9}_{-1.5}$~$\mu$K$^2$, with a 95\% confidence upper limit 4.8~$\mu$K$^2$. 

\paragraph{Q- and W-band Spectra}  In Figure~\ref{fig:results2}, we show power spectrum estimates obtained by analyzing Q- and W-band maps separately. The results are in good agreement with each other and with the combined Q$+$W power spectrum shown previously in Figure~\ref{fig:results1}, providing evidence for a lack of foreground power.

\paragraph{Q$-$W Spectra} We next consider a map obtained by differencing the Q- and W-band maps,
as described in~\S\ref{sec:null_tests}. Since the calibrations for each constituent map included the frequency-dependent thermodynamic correction (see \S\ref{sec:chopper}) appropriate for measuring fluctuations from  a 2.7~K blackbody, this map is insensitive to CMB fluctuations but would be affected by foregrounds. 
Moreover, when power spectra are estimated from this map, their errors do not include sample variance contributions from the CMB.  Thus, analysis of the Q$-$W map probes for foreground contamination more deeply than does checking for consistency between the power spectra of the W- and Q-band individual maps.

Figure~\ref{fig:results3} presents $EE$ and $BB$ estimates from the Q$-$W map.  No significant signal is apparent.  We evaluated the hypothesis that all the $EE$ ($BB$) bandpowers were consistent with zero by summing their equivalent $\chi^2$ statistics; we found 7.4 (8.7) for the $7$ degrees of freedom.  The lack of power in the Q$-$W map suggests that the polarized foregrounds at 40 and 90~GHz are smaller than the 
noise level of the null map ($\approx 90$ $\mu$K-arcmin).  The null result also confirms that the maps from the two frequencies can be reliably combined to produce the Q$+$W spectra in Figure~\ref{fig:results1}.

\paragraph{$EB$ Spectra}
A nonzero $EB$ signal would violate parity invariance and is therefore expected to be absent cosmologically and to be generated only by a limited set of systematic effects, such as an overall error in the detector polarization angles;  of course, foregrounds can also generate an $EB$ signal. (In contrast, $BB$ is parity-invariant and can be generated not only by cosmological effects such as gravity waves and gravitational lensing,  but also by a wider range of systematics.)  Figure~\ref{fig:EBresults} shows estimates of seven bandpowers for the $EB$ cross power spectrum.
The $\chi^2$ to zero is 6.2 for seven degrees of freedom, verifying the expected absence of signal.

\paragraph{The Null Test Suite} We evaluated a set of null tests as described in \S\ref{sec:null_tests}, each based on dividing either the W- or the Q-band data into disjoint subsets to form null maps, which fell into the following five classes: \begin{enumerate} \item Time divisions: difference maps among the three periods, between day/night data, between morning/evening data; and quadrature data.  \item Cryostat divisions:  data from all radiometers in one cryostat versus all the others. \item Spectral divisions: one frequency subband versus the others. \item Housekeeping divisions: differences between the halves of the data with the highest/lowest IF amplifier, LNA, and telescope mirror temperatures. \item Performance divisions:  maps from the polarimetry channels with the best/worst performance in:  $1/f$ noise, ground-synchronous structure, and $I\rightarrow Q$ leakage.\end{enumerate}

In total there were 72 null tests performed on the data: 17 for Q band, 19 for W band, each for $EE$ and $BB$. Figure~\ref{fig:nullsuite} displays the bandpower results from two-band fits for all of the tests; they scatter around zero power.  To assess the performance of the null test suite, for each of the 72 likelihoods we calculated the equivalent $\chi^2$. The distribution of this quantity is plotted in Figure~\ref{fig:nullchis}; it closely 
follows the expected distribution for the hypothesis of zero power throughout the null test suite. The null tests were not fully independent; correlations do not bias the total $\chi^2$ but do increase its variance. The null test suite was studied in full-season simulations and the 
distribution of the resulting total $\chi^2$ in fact agreed well with the expected distribution for a $\chi^2$ with 72 degrees of freedom, indicating that the correlations did not significantly affect the interpretation of this statistic. We computed the total $\chi^2$ for the null suite as 77.8 (the corresponding probability to exceed is 30\%), in confirmation of the hypothesis that the null maps are uncontaminated. The summed $\chi^2$ for subsets of the tests (W- or Q-band alone, just $EE$ or $BB$) revealed no problems, nor did further null tests from the Q$+$W data or from $EB$ bandpowers.

\paragraph{Further Tests}  Several further tests were performed to verify the robustness of the results. We investigated the stability of the data under ring mode removal by analyzing with 3 and 7 modes removed, rather than 5, and by turning off ground-synchronous structure removal. In all three cases, the results changed at less than the $0.1\sigma$ level. We conclude that  the data suffer negligible contamination from scan-synchronous structure, a validation of the efforts to reduce sidelobes described in \S\ref{sec:instrument}. We similarly analyzed the data with a variety of different cut masks, corresponding to various choices of the data selection parameters, and found that the result varied negligibly and within expectations from the statistics of the samples. These expectations were calibrated by following the same procedure on a set of full-season simulations. Finally, we additionally analyzed the data using the alternative responsivity model (model 1 as described in \S\ref{sec:gains}) and found the results changed only by a small amount in agreement with the systematic errors quoted in \S\ref{sec:systematics}.

From the checks and null tests described so far, our results show no evidence
of contamination from either instrumental systematics or CMB foregrounds.  
To make this quantitative, in the following two sections we describe our modeling of possible instrumental systematics and of foreground contamination. 

\section{Systematic Uncertainties}\label{sec:systematics}
Extraction of power spectra from the CAPMAP raw data required the use of several imperfectly determined calibration parameter sets, including 1) the responsivities, 2) the beam sizes, 3) the polarization angles, and 4) the pointing model parameters. We outline in \S\ref{sec:calib} how we determined our best estimates of these parameters, which we call the \emph{baseline parameters}.
Below, we describe how simulations were used to estimate the systematic effects on the power spectrum caused by  measurement and modeling uncertainties for each of these parameter sets, and we then describe a few additional systematic effects that were investigated but found to be negligible.

The analysis codes (for mapmaking and power spectrum estimation) took the calibration parameters as inputs; the results from \S\ref{sec:results} are derived using the baseline parameters.  Of course, the simulations also required calibration parameters to translate CMB realizations into detector timestreams or receiver maps.  Our usual approach was to generate hundreds of simulations using the baseline parameters, and then to analyze each one 
using perturbed estimates of the calibration parameters in the parameter set under study. These new estimates were drawn from distributions consistent with measurement and/or modeling uncertainties.
In addition, each simulation was also 
subject to the standard analysis. The two analyses of each simulation were compared, and the shifts in the likelihood peaks for each of the 14 $\ell$ bands were recorded. 

For every  parameter set studied, the mean peak shift was consistent with zero.
The standard deviations of the peak shift distributions were used to estimate the systematic uncertainties,
which are listed in Table~\ref{tab:sys_effects} for the Q$+$W $EE$ bands.  We also estimated systematic errors for  W- and Q-band separately, finding similar results. The table does not list the $BB$ systematic effect levels; they are comparable, in comparison with the statistical error bars, to the $EE$ levels.

\subsection{Responsivity}
The time dependence of the responsivity for each polarimeter was modeled as described in \S\ref{sec:gains}.  We considered two models and chose the simpler one, which ignored short-timescale variations.  To evaluate the impact on the power spectrum of neglecting the faster variations, we generated 100 full-season simulations and, for each, compared results from the standard analysis to results from analysis using the alternate responsivity model, which included short-timescale variations.
The resulting systematic error  estimate for each  bandpower was typically 10--20\% of the statistical error, increasing the overall error in the worst case by 3\% after adding in quadrature. 
 
To bound the effects of these inconstant responsivities, we also analyzed the simulations under the unreasonable assumption that the responsivities were constant for the duration of the season.
In this case---in which we did not account for the period II anomaly, the time-varying opacity, or even for the well-understood gradual responsivity decrease over the course of the season---the derived systematic error only doubled.

The Q$+$W result was also subject to an additional systematic error (not affecting Q- or W-band individually) due to the relative calibration uncertainty between the two frequencies, which was dominated by the uncertainty on the W-band beam efficiency discussed in \S\ref{sec:chopper}. This was studied in receiver map simulations by varying the relative calibration within the $\approx\pm5\%$ distribution expected from the optical simulations and was found to be subdominant in all cases to the responsivity model uncertainty. The first column of Table~\ref{tab:sys_effects} reports the two contributions added in quadrature.

These systematic error estimates were perhaps smaller than might have been predicted in light of the 10\%-level effects ignored in the baseline responsivity model.  Two features of CAPMAP provided immunity:  the large number of channels, and the fact that the entire sky was surveyed every six hours, which diminished the effect of short-timescale variations.
We note that though these systematic errors are small relative to our statistical errors, they are not small compared to predictions of the primordial BB spectrum; fortunately this systematic effect would be greatly reduced with successful  temperature control to stabilize the responsivities.

\subsection{Beams}
Observations of Jupiter were used to find the beamsize for every receiver to a few percent, as described in \S\ref{sec:beams}.  Each W-band (Q-band) beamsize was within 5--6\% of the W-band (Q-band) mean.
We took advantage of the good agreement by treating all the W-band (Q-band) channels as having the same beamsize, which facilitated combination of the data from separate polarimeters in the mapmaking stage.  We evaluated the systematic impact of this approximation by generating simulations in which 
each receiver map resulted from convolution with the actual measured beamsize of that receiver, and then subjecting the maps to the standard analysis. 
The resulting systematic error was not the dominant effect in any $\ell$ band, validating our simplified treatment of the beams in the pipeline.

We also investigated a possible uncertainty from ignoring the $\lesssim 8\%$ elongations of the beams.  We numerically calculated the window function (before the complication of scanning) for the worst elongation, and then differenced it from the  symmetric one used in the analysis.  Properly folding this difference with the concordance model $EE$ power spectrum showed that, for all bands, the corresponding correction was a small fraction of the beam uncertainty discussed above so that it could be safely neglected.

Finally, we used a similar method to evaluate the validity of our treatment of the beams as Gaussian, in view of the sidelobes described in  \S\ref{sec:chopper}. We modeled the sidelobes by adding a second Gaussian to the main one, calculated a simplified window function, and again found this effect negligible compared to the beam uncertainty estimated above. 

\subsection{Pointing}
Results from the pointing solution are described in \S\ref{sec:pointing}. The largest uncertainty was in determining  the global pointing:  the position of the center of the survey region (nominally the NCP). The associated systematic effect was evaluated from 500 receiver map simulations, each analyzed with an estimate for the azimuth and elevation of the survey center drawn from Gaussian distributions centered at the NCP with widths $18\arcsec$ and $29\arcsec$ respectively.  The resulting systematic error was not the dominant effect in any $\ell$ band. The offsets of the polarimeters from the array center were better constrained by measurements than was the global pointing; simulations confirmed that they led to a negligible additional contribution to the power spectrum error. 

\subsection{Polarization Angles}
The detector polarization angles were measured with a typical error of $1.5^\circ$ as described in \S\ref{sec:polznang}.
We evaluated the systematic errors reported in Table~\ref{tab:sys_effects} with 500 receiver map simulations, analyzing each one with a set of detector angles drawn from Gaussian distributions with $\sigma=1.5^\circ$ about their baseline values. 
We also considered the effect of a global bias in the detector angles, which would rotate $E$-modes into $B$-modes and generate non-zero $EB$ power. The few-degree precision of our measurements of each detector polarization angle limited the possible magnitude of a global bias sufficiently to make its impact negligible.

\subsection{Other Effects}
We explored a few other potential sources of systematic uncertainty and found them to be smaller than the others. These have not been included in Table~\ref{tab:sys_effects}; instead, we 
indicate here their small impacts by giving the uncertainty estimated using a two-band analysis of simulations for Q$+$W. We report the uncertainties as pairs:  [$X$,$Y$], in which $X$ is a multiplier to the concordance model for $EE$ and $Y$ is the uncertainty for $BB$ in $\mu$K$^2$.  

As discussed in \S\ref{sec:opacity}, there was an additional systematic uncertainty from the opacity determination arising from unmodeled responsivity variations in the total power channels used in its measurement.  This uncertainty was estimated by analyzing full-season simulations with an alternative opacity model relying on different total power channels with significantly different responsivity variations.  The results were [0.005, 0.03].

We also studied the effects of  the $I\rightarrow Q$ leakage described in \S\ref{sec:IQleak}. Worst-case estimates for the effect of the monopole and dipole terms predicted that they would contribute at most 2\% to measurements of the $EE$ power spectrum \citep{jeff_thesis}. We simulated the effects of the monopole term by adding appropriate amounts of $TT$ to the polarization maps and found the uncertainties to be  [0.003, 0.04].  The $EE$ effect was consistent with zero, with an uncertainty that was nearly a factor of 10 smaller than the 2\% estimate because of suppression from the ring scan strategy: each spot on the sky was measured with many parallactic angles.  Quadrupole leakage is not suppressed by scanning or sky rotation, but the CAPMAP optics were specifically optimized to keep the quadrupole small.  We estimated the quadrupole contribution to the measured $EE$ spectrum to be less than 0.1\%, smaller than the monopole systematic.

Pipeline 2 was subject to an additional systematic error due to the assumption of LST-invariant noise (\S\ref{sec:mapmaking}). The effect of ignoring the $\approx10\%$ variation of the noise with LST was studied in receiver map simulations, which showed no detectable bias. The  uncertainties on $EE$ and $BB$ were [0.004, 0.02].

As mentioned in \S\ref{sec:data_selection}, some of the polarimeters suffered enough $1/f$ noise to perturb slightly the distributions of their $\chi^2$ statistics from the five-ring-mode fits performed on each cycle during the data selection process, resulting in a high grand $\chi^2$. The analysis pipelines assumed white noise (\S\ref{sec:mapmaking}).  We evaluated the systematic impact of this assumption by generating full-season simulations including $1/f$ noise (using the method of \citet{one_over_f}) in each polarimeter consistent with measured noise spectra, and subjecting them to the standard analysis. We found a negligible bias, and the systematic uncertainty was $[0.003, 0.05]$. Finally, we note that the parameter-division grand $\chi^2$ null test (equivalent to differencing maps made from channels with high and low levels of $1/f$ noise)
passed at the $1\sigma$ level for both $EE$ and $BB$ and in both frequency channels (Fig.~\ref{fig:nullsuite}).

\subsection{Systematics Summary}

\begin{deluxetable}{lccccr}
\tablecaption{\label{tab:sys_effects}Q$+$W Systematic Effects}
\tablehead{
 \colhead{Band} & \colhead{Responsivity} & \colhead{Beams} & \colhead{Pointing} & \colhead{Angles} & \colhead{Total}}
\startdata
1 & $0.6$ & $0.2$ & $0.3$ & $0.2$ & $0.7$ ($0.15\sigma$)\\
2 & $1.4$ & $0.5$ & $0.7$ & $0.3$ & $1.7$ ($0.18\sigma$)\\
3 & $1.8$ & $1.2$ & $1.3$ & $0.9$ & $2.7$ ($0.21\sigma$) \\
4 & $2.5$ & $1.7$ & $1.9$ & $2.0$ & $4.1$ ($0.27\sigma$)\\
5 & $4.8$ & $3.4$ & $3.9$ & $3.7$ & $8.0$ ($0.37\sigma$) \\
6 & $6.6$ & $4.0$ & $4.5$ & $4.8$ & $10.1$ ($0.29\sigma$)\\
7 & $5.3$ & $3.1$ & $3.7$ & $3.0$ & $7.8$ ($0.29\sigma$)
\enddata

\tablecomments{Systematic error levels are given for all 7 Q$+$W $EE$ power spectrum evaluation bands; the errors for W- and Q-band separately or for $BB$ are comparable relative to the respective statistical errors. All errors are quoted as flat bandpowers in $\mu$K$^2$ and are calculated as described in \S\ref{sec:systematics}. The rightmost column gives the total systematic error added in quadrature; this quantity is to be added in quadrature with the statistical errors given in Table~\ref{tab:results}. Shown in parentheses is this same quantity divided by the half-width of the $1\sigma$ confidence interval, which allows the systematic uncertainty to be compared with the size of the statistical error bar. Not reflected in these numbers is an overall $\approx20\%$ calibration uncertainty affecting the results and the error bars equally.}
\end{deluxetable}

The rightmost column of Table~\ref{tab:sys_effects} gives the total systematic error added in quadrature, first in $\mu$K$^2$ and then in units of  the statistical error. The largest systematic effect relative to the statistical error is for the $EE$ 1401--1700 band. There the systematic error is 37\% of the statistical error, increasing the total error for that band by 6.6\%. 

We have also calculated the total systematic effect from all the contributions considered above (by adding the errors in quadrature) for the two-band fits as an overall summary: the result is 2\% of the concordance model for $EE$ and $0.13~\mu$K$^2$ for $BB$, both in the 2--3000 band.

In addition to these effects, there is an overall multiplicative uncertainty of 
20\% (16\%) in W-band (Q-band) on both the power spectrum results and their error bars, after adding in quadrature the 9.4\% (7.6\%) absolute uncertainty on the responsivity calibration and the $2\%$ uncertainty on the opacity determination (and taking the appropriate square since the results are expressed in units of $\mu$K$^2$).
The overall calibration uncertainties on the Q$+$W results varied with $\ell$ and were studied in simulations. At low $\ell$, Q- and W-band contributed equal weight to the result, producing a net calibration uncertainty of 18\%. At high $\ell$, only W-band contributed, and the calibration uncertainty is 20\%.
These errors do not affect the detection significance.

\section{Foregrounds}\label{sec:foregrounds}

From results presented in \S\ref{sec:results}, we find no evidence for significant foreground contamination in our data: no power in the Q$-$W map and no significant $BB$ power in any of our maps.

We also tested the Q- and W-band maps for possible point source contamination by excising pixels in which the measured signal exceeded the mean by $3.5 \sigma$ (approximately 70 $\mu$K at the $3.4\arcmin$ resolution used for this study).
There were 4 (2) such pixels in the Q-band (W-band) map; masking these pixels produced a negligible change in the multi-band power spectra.

Such tests have limited statistical power, however, so we supplemented them with simulations.  Here we describe our methodology \citep{keith_thesis} and results for both point sources and diffuse galactic contributions. 

Using the full-sky map of synchrotron radiation provided by \citet{haslam}, we extrapolated with $\beta = -3$ to produce synchrotron intensity templates at 40 and 90~GHz. Similarly, following the
method of \citet{finkbeiner}, which models thermal dust as two
distinct populations with spectral indices of 1.7 and 2.7, dust
intensity templates were produced at the two frequencies. 
Only the Q-band power from these templates needed to be scaled down by a factor of about four to match that reported by WMAP \citep{kogut} within the CAPMAP survey region.

The polarization angle at each point on the sky was chosen
following  \citet{giardino}, and polarization
levels of 10\% and 5\% were taken as  conservative estimates for
synchrotron and thermal dust, respectively.  We took the two polarizations to be aligned, as a worst but not implausible case.

The WMAP point source catalog \citep{hinshaw} is currently the only one that includes the
NCP (at galactic latitude $27.1^\circ$). It contains no sources within the CAPMAP survey region, to thresholds of
roughly 300~$\mu$K and 500~$\mu$K in Q- and W-band, respectively.

Point sources bright in microwaves and outside the galactic
plane are dominated by Active Galactic Nuclei, the vast majority of
which are blazars (Giommi \& Colafrancesco, 2006):  flat spectrum ($\alpha \approx -0.25$, where $S_\nu \sim
\nu^\alpha$) and weakly ($<10\%$) polarized. A recent comprehensive blazar survey \citep{healey}
found no sources with flux at 4.8~GHz greater than 65~mJy ($\sim200$~$\mu$K for
a W-band receiver, assuming $\alpha = -0.25$) within the CAPMAP survey region.

To proceed, we simulated the effects of fainter sources, guided by the well-measured statistics of the blazar population (see e.g.\ \citet{giommi}, \citet{giommi2}, \citet{deZotti}).
We assumed a polarization level of 3\%.
Our simulations agree well with \citet{tucci}, who estimate power spectra from extragalactic radio sources, taking into account measured source polarization distributions at 1.4~GHz, and including both steep and flat-spectrum sources. 

For each of 100 simulations, a foreground realization was added to a map with a CMB and noise realization matching the final data set. For W-band, the mean change in power was never more than 1~$\mu$K$^2$ in any band. In Q-band, the effect was at the level of $5\%$ of concordance; we find, for example, at $\ell=1000$ about 1~$\mu$K$^2$ from faint point sources and 0.5~$\mu$K$^2$ from synchrotron radiation. We conclude that our W-band results are likely foreground free while those in Q-band could be contaminated at the level of $\approx 5\%$.  However, the simulations show that our main Q$+$W results should be contaminated at less than $2\%$.

\section{Pipeline Complementarity and Comparisons with Our Previous Work }\label{sec:comparisons}
In this section we summarize the roles of the two pipelines.  We follow with summary comparisons to our previous work.
 
\subsection{Complementarity of Pipelines}
 Our work has benefited from having two pipelines in a number of ways. The agreement of the results from the two pipelines is strong evidence that the CMB results are robust.  The two pipelines traded simulations to verify their code implementations. As described in \S\ref{sec:data_selection}, each initially developed independent data selection methods, though the cut variables developed by pipeline 1 were discovered to be the more effective and were eventually adopted by both pipelines; it was also discovered that it was necessary to include the longer timescale explored by pipeline 2.

Having two pipelines allowed us to subject the data and its analysis to a wide battery of tests; neither pipeline had to implement every test.
 
Pipeline 1 was in a sense a conventional implementation.  It was capable of producing likelihood estimators with arbitrary noise realizations and without significant approximation.  However, when the pixel count exceeded $\approx 10^4$, the processing time became prohibitive.  The majority of studies with this pipeline were done with about 2500 identically sized $3.4\arcmin$ pixels, the size of the W-band beam. These studies included: the evaluation and optimization of possible scan strategies; development of the null-test suite, the Q$-$W and $EB$ null tests, development and study of the selection criteria; and preliminary studies of the effects of $1/f$ noise in the timestreams.

\begin{figure*}
\centering
\includegraphics[width=\textwidth]{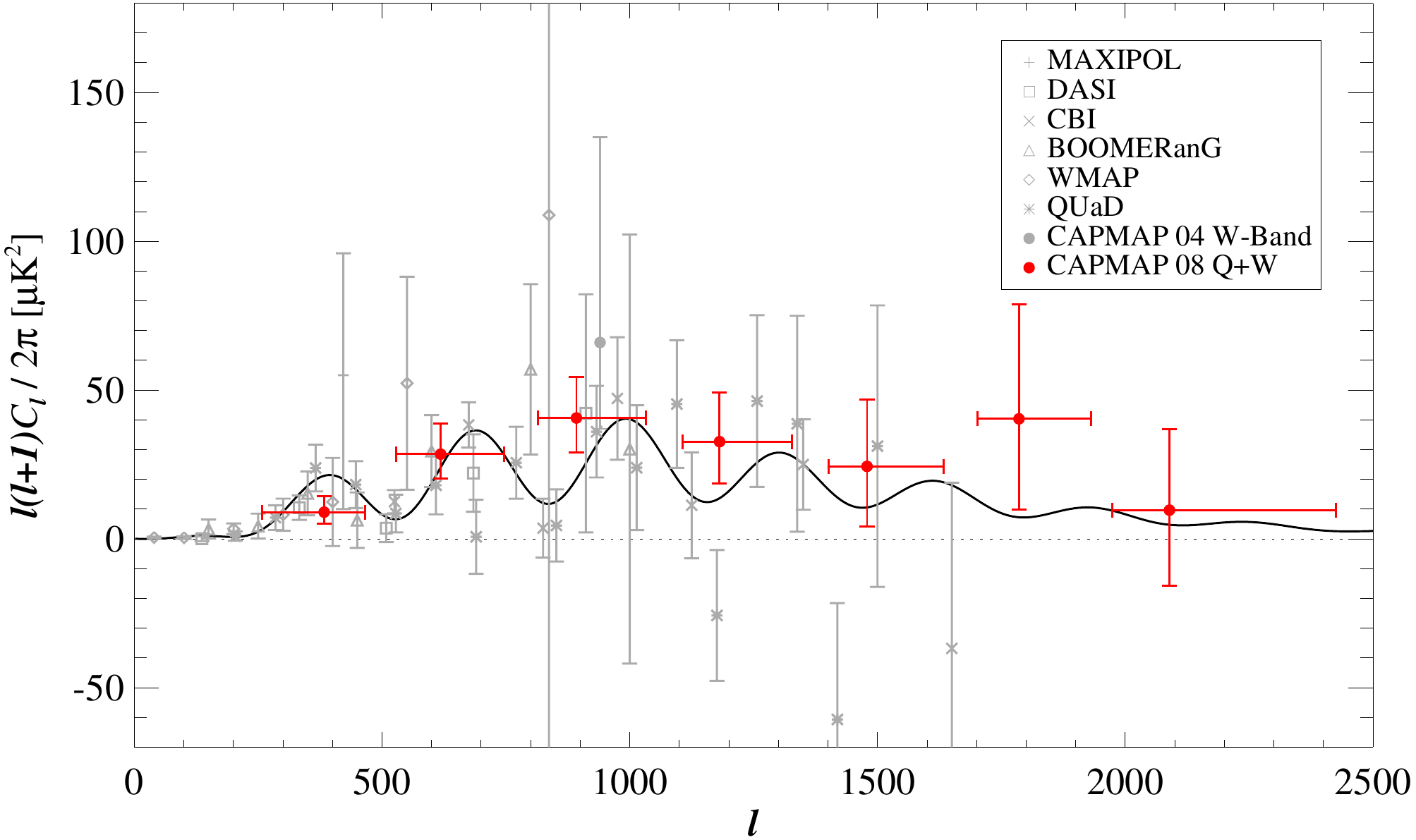}
\caption{\label{fig:results}Compilation of all measurements to date of the CMB $EE$ power spectrum. Values are taken from the present work (pipeline 2) and from  MAXIPOL \citep{maxipol}, DASI \citep{dasi}, CBI \citep{cbi}, BOOMERanG \citep{boomerang}, WMAP \citep{wmappol}, QUaD \citep{quad}, and the CAPMAP first season \citep{Barkats:2004he}. The error bars show statistical errors only in all cases.}
\end{figure*}

Pipeline 2 was more novel, exploiting the speed advantages of the half-Fourier technique.  The calculation of the likelihood from the data vector at one point in parameter space needed less than 1 CPU second.  This enabled the fast simulation and analysis of a variety of systematic effects with $1.2\arcmin$ pixels, including those reported in Table~\ref{tab:sys_effects}; in particular, evaluation of the impact of responsivity uncertainties required hundreds of full-season simulations from pipeline 2.  Other studies with this pipeline included evaluation of the null test suite with $1.2\arcmin$ pixels with many different data selection schemes, simulation of the null test suite, and study of the impact of $1/f$ noise with full-season simulations.  Pipeline 2 required the noise for each polarimeter to be approximately independent of LST. Although the noise condition was satisfied for the full-season data set and was also adequately satisfied for the three periods, this limitation prevented pipeline 2 from performing such tests as the day/night null test and the foreground study in which several possible point sources were excised.

Pipeline 2 ran routinely with $1.2\arcmin$ pixels and determined 14-band conditional likelihoods in 48 CPU hours;  pipeline 1, running with $1.7\arcmin$ pixels, took about 175 CPU hours for the same task ($1.7\arcmin$ pixels were small enough, given the noise levels). The final results were produced with the MCMC method; the implementations were different, with pipeline 1 taking about 10000 CPU hours and pipeline 2 about 1500 CPU hours.
 
\subsection{Comparisons with Our Previous Work}
Our first publication reported a $2\sigma$ detection over one large band obtained from one season's worth of data with 4 W-band radiometers. This work is a significant improvement over that result, most importantly because we operated for more time with a greatly expanded array and introduced a second frequency.

To capitalize on the improved statistics, for this work we treated systematic effects in much more detail, including incorporation of the beam efficiency effect that we had taken to be negligible in the past; in fact the magnitude of that effect was equal to the calibration error we reported in \citet{Barkats:2004he}. Several improvements reduced our susceptibility to systematic effects.
One was the improved AR coatings on the lenses (\S\ref{sec:instrument} and \S\ref{sec:calib}), which greatly reduced the polarimeters' sensitivity to unpolarized point sources (subdominant to the expected contribution from polarized point sources).
Another was the reduction in telescope sidelobes (\S\ref{sec:instrument}), which  made ground-synchronous structure (\S\ref{sec:data_selection}) negligible, eliminating concerns that variations in such residual scan-synchronous signal might contaminate the data.  Finally, the  highly redundant and symmetric ring scan strategy provided another layer of immunity. 

\section{Final Results and Conclusions}\label{sec:conclusions}

\begin{deluxetable}{clll}
\tablewidth{2.75in}
\tablecaption{\label{tab:results}Q$+$W Power Spectrum Results}
\tablehead{\colhead{Band} & \colhead{$\ell$} & \colhead{$EE$ ($\mu$K$^2$)} & \colhead{$BB$ ($\mu$K$^2$)}} \startdata
$1$ & $383^{+83}_{-125}$ & $9.0^{+5.3}_{-3.8}$& $-1.1^{+2.1}_{-1.3}$\\
$2$ & $618^{+128}_{-89}$ & $28.5^{+10.3}_{-8.3}$& $2.6^{+5.1}_{-3.6}$\\
$3$ & $892^{+139}_{-77}$ & $40.6^{+13.7}_{-11.5}$& $7.7^{+9.0}_{-7.1}$\\
$4$ & $1181^{+145}_{-74}$ & $32.7^{+16.4}_{-14.1}$& $-9.2^{+10.3}_{-8.2}$\\
$5$ & $1478^{+154}_{-77}$ & $24.4^{+22.4}_{-20.2}$& $21.4^{+23.1}_{-19.1}$\\
$6$ & $1785^{+145}_{-83}$ & $40.4^{+38.4}_{-30.5}$& $22.8^{+33.5}_{-27.3}$\\
$7$ & $2089^{+336}_{-116}$ & $9.7^{+27.2}_{-25.3}$& $-29.3^{+31.6}_{-25.0}$\\
\enddata
\tablecomments{All results are in $\mu$K$^2$ and are expressed as maximum likelihood flat bandpowers with 68\% confidence intervals. The effective multipoles are listed for the $EE$ bands; those for $BB$ differ only slightly. The errors given are statistical errors only; the systematic errors are discussed in \S\ref{sec:systematics}.}
\end{deluxetable}

\begin{figure}
\centering
\includegraphics[width=3.39375in]{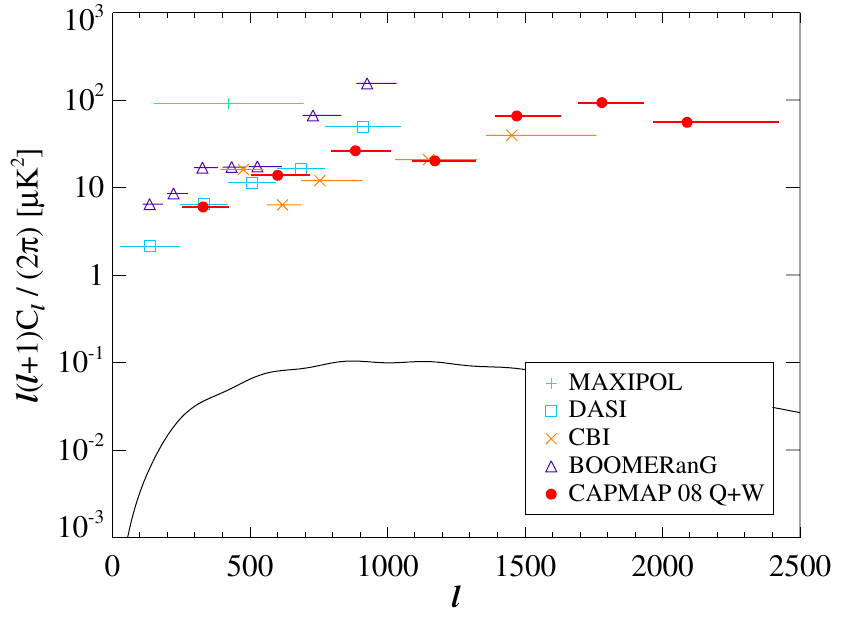}
\caption{\label{fig:bblimits}95\% confidence upper limits on the $BB$ power spectrum from CAPMAP and other recent experiments (as cited in Fig.~\ref{fig:results}). We show limits (neglecting calibration uncertainties) from all published experiments that report limits on $BB$ power.  Limits were calculated by integrating only the part of the likelihood corresponding to $C^{BB}_\ell \geq 0$. For MAXIPOL and DASI, published upper limit values were used, while for CBI and BOOMERanG, upper limits were calculated from offset lognormal approximations \citep{bjkrad} to the likelihood functions, using published parameters. QUaD does not report upper limits. The expected signal from gravitational lensing (smooth curve) shows the next target for fine-scale CMB polarization.}
\end{figure}

Figure~\ref{fig:results} shows our Q$+$W $EE$ results together with those of all other experiments that have reported $EE$ measurements. Both $EE$ and $BB$ bandpowers are listed in Table~\ref{tab:results}. The excellent agreement between the pipelines means that it matters little which set of results we report; we have arbitrarily chosen those of pipeline 2.

We detect polarization at $11\sigma$, and our highest $\ell$ bandpowers, while not of sufficient precision for a detection at the expected level, nevertheless 
extend the community data set into these scales. Our results, like all those already published, are in good agreement with the concordance model. Figure~\ref{fig:bblimits} presents our $BB$ results, interpreted as 95\% CL exclusions, again in comparison to previous work.  While no experiment is yet sufficiently close to detection of gravitational lensing, our sensitivity is among the best.

We mention again our result on the Q$-$W  map (Fig.~\ref{fig:results3}). Unlike the suite of 72 null tests described in \S\ref{sec:results}, this one did not have to pass; that it did pass constitutes an additional science result---we find that maps from both frequencies are consistent with the same underlying CMB realization and hence limit foreground contamination.

Our $Q$/$U$ coverage was very uniform, owing to the ring scan strategy, so that our $EE$ and $BB$ bandpowers were largely uncorrelated.  The scan strategy provided other benefits, including continuous monitoring of the opacity and data quality. We exploited the symmetry of the survey region to explore the speed advantages of analysis in a half-Fourier space. Our two complementary pipelines were fully tested with end-to-end simulations, and a variety of systematic uncertainties were estimated from detailed simulations.  

We emphasize here that our approach was conservative: our selection criteria, in particular, were (in part) based on  the performance of the suite of null tests.  We note that while the null tests might have passed or failed depending on choices of cuts, the final results themselves were quite insensitive to cutting parameters.  The situation was the same with ground-synchronous structure (\S\ref{sec:data_selection}):  we projected out such a mode in the analysis, and reported results with the associated loss in sensitivity,  but in fact not doing so made no significant change to the recovered central values. Similarly, we projected out five modes associated with the ring scan, but analyzing with only three modes removed left the results statistically unchanged (but with smaller errors).

It might seem surprising that the polarization results from CAPMAP, which took data for just a few months and observed from not a particularly ideal location, compare to those from other experiments operating with comparable numbers of detectors and observing for longer times from sites regarded as the best on the planet.  There are a few factors that contributed to this: our noise was sufficiently well understood that we were able to use auto- (rather than only cross-) correlation power spectra; the stability of our residual ground-synchronous structure allowed us to remove ground contamination with a smaller number of modes than experiments that use field differencing; and the 7~m  telescope allowed us to center our sensitivity in multipole space where the expected signal, for both $E$- and $B$-modes, is maximal.

The next target at fine angular scales is the detection of $BB$ power from lensing; reaching this goal will be a challenge, requiring more detectors and even better immunity from systematic uncertainty.
 
\acknowledgements
We thank Dale Fixsen for convincing us to try a ring scan strategy, and Chris Hirata for arguing the virtues of the half-Fourier map space to us. We thank John Carlstrom, Tom Crawford, Joe Fowler, Wayne Hu, Norman Jarosik, Steve Meyer, Lyman Page, and David Spergel for many helpful conversations; Rob Lucy, Eugenia Stefanescu, and Michelle Yeh for help with the data collection; and Glenn Atkinson, Bill Dix, Ted Griffiths, Mike Peloso, Elizabeth Pod, Fukun Tang, and Laszlo Varga for their contributions to the design and construction of the experiment.
Many thanks to Neelesh Arora, Valeri Galtsev, and Vinod Gupta for computing support, and to Julian Borrill for managing the CMB analysis group at the NERSC cluster. We are grateful to Lucent Technologies for the use of the 7~m antenna and to Tod Sizer, Bob Wilson, and Greg Wright for their support and assistance.
We gratefully acknowledge support from NSF grants PHY-9984440 and PHY-0355328 and Princeton University; and the Kavli Institute for Cosmological Physics, funded by the Kavli Foundation and NSF grant PHY-0551142.
We acknowledge the use of the Legacy Archive for Microwave Background Data Analysis (LAMBDA). Support for LAMBDA is provided by the NASA Office of Space Science. Some of the results in this paper have been derived using the HEALPix \citep{HEALPix} package.

\end{document}